\documentclass[aps,floats,showpacs,showkeys,preprint]{revtex4}
\usepackage{amsmath}
\usepackage{amssymb}
\usepackage{amsthm}
\usepackage{graphicx}
\usepackage{dcolumn}
\newcommand{\be}{\begin{equation}}
\newcommand{\ee}{\end{equation}}
\newcommand{\ba}{\begin{eqnarray}}
\newcommand{\ea}{\end{eqnarray}}

\begin{document}
\title{Inverse melting in lattice-gas models}
\author{Santi Prestipino}
\email{Santi.Prestipino@unime.it}
\affiliation{Universit\`a degli Studi di Messina, Dipartimento di Fisica, Contrada Papardo, 98166 Messina, Italy}
\date{\today}

\begin{abstract} ~~Inverse melting is the phenomenon, observed in both
Helium isotopes, by which a crystal melts when cooled at constant pressure.
I investigate discrete-space analogs of inverse melting by means of two
instances of a triangular-lattice-gas system endowed with a soft-core
repulsion and a short-ranged attraction. To reconstruct the phase diagram,
I use both transfer-matrix and Monte Carlo methods, as well as
low-temperature series expansions.
In one case, a phase behavior reminiscent of Helium emerges, with a
loose-packed phase (which is solid-like for low temperatures and
liquid-like for high temperatures) extending down to zero temperature
for low pressures and the possibility of melting the close-packed solid
by isobaric cooling.
At variance with previous model studies of inverse melting, the driving
mechanism of the present phenomenon is mainly geometrical, related to the
larger free-energy cost of a ``vacancy'' in the loose-packed solid than in
the close-packed one.
\end{abstract}



\maketitle
\thispagestyle{empty}

%
%
\section{Introduction}
\setcounter{page}{1}
\renewcommand{\theequation}{1.\arabic{equation}}

Inverse melting (IM) is hardly mentioned in a catalog of Helium oddities.
Yet, Helium provides the only example of an elemental solid that can be
melted isobarically by {\it lowering} the temperature, although this
only occurs in a small range of pressures.
This striking phenomenon, which is still poorly understood, is evidenced
in the low-temperature phase diagrams of both $^3$He and $^4$He by a
decreasing profile of the solid-liquid coexistence pressure $P_{\rm cox}$
as a function of temperature $T$~\cite{Balibar}.
Another kind of IM is provided by P4MP1 polymer solutions~\cite{Rastogi},
where a tetragonal crystal undergoes amorphization on cooling, accompanied
by heat release.

The existence of IM fights against common sense
but by no means disproves thermodynamics~\cite{Tammann,Stillinger}.
The slope of $P_{\rm cox}(T)$ is ruled by the Clausius-Clapeyron equation,
\be
\frac{{\rm d}P_{\rm cox}}{{\rm d}T}=\frac{S_2-S_1}{V_2-V_1}\,,
\label{1-1}
\ee
where $V_1,V_2$ and $S_1,S_2$ are the volume and entropy values of the
coexisting phases.
Nothing prevents a phase transition to occur, at fixed external pressure,
from a low-temperature phase 1 to a high-temperature phase 2 with
both volume contraction and positive latent heat, hence with negative 
$\Delta S/\Delta V$ (this is the case, for instance, of ordinary ice
and water). What is peculiar to Helium IM is that
the more compact and more entropic phase is solid rather than liquid.
The case of P4MP1 is different: Here $\Delta S/\Delta V>0$, but
the solid still lies on the high-temperature side of the transition.

Generally speaking, IM requires some microscopic mechanism
by which spatially-confined particles can nonetheless have more entropy
than the coexisting liquid. In $^3$He, the nuclear spins are more
coupled ({\it i.e.}, less free to orient independently of each
other) in the liquid than in the solid phase.
Inverse melting of P4MP1 is explained by a larger amount of polymer
conformations in the crystal than in the amorphous state, due to a more
open crystal structure.

Pursuing the analogy with the above systems, the few models of IM
proposed so far have invariably focussed on systems of particles
with internal degrees of freedom~\cite{Feeney,Schupper}.
When the single-particle states are taken to be more degenerate in the
ordered than in the disordered phase, IM can occur.
In practice, one needs a very fine tuning of the single-particle
spectrum to obtain a realistic IM scenario.

The purpose of this paper is to show that IM -- or an analog of it --
is also observed in systems that are hosted on a lattice.
The model presented here is a classical lattice-gas system which, with
no other ingredients than the radial dependence of its pair-interaction
strenght, exhibits a phase behavior that is reminiscent of Helium.
In particular, 1) the system can exist in a loose-packed phase down to
zero temperature for low enough pressures; and 2) in a range of pressures,
the close-packed solid melts into a less dense structure upon cooling.
This phase is crystalline at very low temperatures, but can be a liquid
at higher temperatures.
Besides the similarities, however, the present IM has conceivably little
to do with Helium: While IM of the latter has a quantum-mechanical origin,
the phenomenon herein discussed heavily relies on the steric constraints
being determined by the interplay between lattice geometry and interaction.
Specifically, it deals with the different free-energy cost, at low
temperature, of vacancy-like excitations in the loose-packed solid and in
the close-packed one. In fact, any lattice potential that is provided with
an extended hard-core repulsion, a shoulder, and a thin attractive well
would prove adequate to bend downward $P_{\rm cox}(T)$ at $T=0$.

The rest of the paper is organized as follows: In Section 2, after introducing
a class of lattice gases that could possibly share some features in common
with Helium, I outline an analytic method for checking the existence of IM.
Then, in Section 3, I describe the numerical techniques that are used
to work out the phase diagram of such a lattice gas.
Results are presented in Section 4, where a comparison is made between
two similar case studies, in an attempt to unveil general rules of
behavior. Some further remarks and conclusions are given in Section 5.

%
%
\section{Model}
\renewcommand{\theequation}{2.\arabic{equation}}
\setcounter{equation}{0}

Classical lattice gases provide caricatural descriptions of real fluids
that, while broadly preserving the topology of the phase diagram, are
by far much simpler to study both analytically and
numerically~\cite{Runnels1,Orban1,Poland,Prestipino1}.
If the aim is finding a lattice-gas system with a Helium-type phase
diagram, but with no ambition to model a real substance, working in
two dimensions is by no means restrictive, in fact it is convenient
for a twofold reason:
The transfer-matrix method, which is the most powerful technique for
reconstructing numerically the phase diagram of a lattice gas,
is a viable tool only in two dimensions.
Moreover, with an eye to making use also of low-temperature expansions
for the analysis, enumeration of ground-state defects/excitations is
much easier in two than in three dimensions.

The type of models here considered is a lattice-gas system hosted on
the triangular lattice.
A given lattice site can be either occupied (by a single particle)
or empty. Calling $c_i=0,1$ the occupation number of site $i$,
the lattice-gas Hamiltonian reads $\sum_{i<j}v(|i-j|)c_ic_j$,
with a pair potential $v$ depending on site-site separation only.
In the very first place, this potential is asked to meet two conditions:
1) It must allow for the existence of two disordered phases, ``vapor''
and ``liquid''; 2) two different crystalline phases should be stable at
$T=0$, say ``solid A'' at high pressures and ``solid B'' at low pressures.
However, the real challenge is to find a model where ``solid B'' and
``liquid'' are actually the same phase or, at least, have similar
thermodynamic properties along the transition line.

I have shown elsewhere~\cite{Prestipino1} that an extended hard core in
the potential, when associated with an attractive tail, brings forth two
distinct fluid phases into the system.
Likewise, to observe a pressure-driven solid-solid transition at $T=0$,
$v(r)$ must have a shoulder before the minimum~\cite{Orban1,Wilding,Quigley};
with this trick, the minimum-energy configuration (which is dominant at low
temperatures and pressures) is kept distinct from the close-packed
configuration (which is preferred at high pressures).
Yet, these two requirements alone do not imply a unique potential,
since the extent of the hard core ({\it i.e.}, the particle diameter)
as well as the height and width of the shoulder relative to the well
remain at will. The following potentials, defining the LG34 and LG56
models, are just two of a host of choices:
\be
v_{34}(r)=\left\{
\begin{array}{rl}
+\infty\,, & \,\,\,{\rm for}\,\,r\leq r_2 \\
0\,, & \,\,\,{\rm for}\,\,r=r_3 \\
-\epsilon\,, & \,\,\,{\rm for}\,\,r=r_4 \\
0\,, & \,\,\,{\rm for}\,\,r\geq r_5
\end{array}
\right.
\,\,\,\,\,\,\,\,\,\,{\rm and}\,\,\,\,\,\,\,\,\,\,
v_{56}(r)=\left\{
\begin{array}{rl}
+\infty\,, & \,\,\,{\rm for}\,\,r\leq r_4 \\
0\,, & \,\,\,{\rm for}\,\,r=r_5 \\
-\epsilon\,, & \,\,\,{\rm for}\,\,r=r_6 \\
0\,, & \,\,\,{\rm for}\,\,r\geq r_7\,,
\end{array}
\right.
\label{2-1}
\ee
where $r_n$ is the distance between a pair of $n$-th neighbor lattice
sites and $\epsilon>0$ sets the temperature scale.
A pictorial description of these models can be found in Fig.\,1.

To determine whether an IM is possible or definitely excluded in a lattice
gas with two distinct solid ground states, the low-temperature profile of
the solid-solid $P_{\rm cox}(T)$ is constructed, looking for a downward
bending near $T=0$.
However, a negative slope is not sufficient evidence of a genuine IM
{\em unless} a continuous path exists from solid B to liquid, a question
that can only be settled numerically, by {\it e.g.} the transfer-matrix
method.

I recall that, for given temperature $T$ and chemical potential $\mu$,
the equilibrium state of a system with fixed volume $V$ is the one
minimizing the generalized thermodynamic potential
$\tilde{\Omega}=E-TS(E,V,N)-\mu N$ as a function of energy $E$ and
particle number $N$, $S(E,V,N)$ being the system entropy
function~\cite{Prestipino2}.
The minimum $\tilde{\Omega}$ defines the system grand potential
$\Omega$, whereas $-\Omega/V$ is the equilibrium pressure $P$
expressed in terms of $T$ and $\mu$.
At $T=0$, the eligible configurations of a lattice gas are only a few,
each continuously linked with a possible system phase at $T>0$.
$T=0$ transitions occur in coincidence with any jump in the values
of $E$ and $N$ at the point of minimum of $\tilde{\Omega}$.

In the LG56 model, for example, only three states are involved at $T=0$,
namely two triangular crystals (solid A and solid B) and the vacuum (the
$T=0$ vapor) -- other solid phases, with rectangular rather than triangular
symmetry, are excluded.
With a slight abuse of terminology, the same three phases are hereafter
termed solid, liquid, and vapor (in other words, I use ``solid B'' and
``liquid'' interchangeably as synonyms of a low-temperature loose-packed
phase).
In the liquid phase, the number density $\rho$ takes values that are
intermediate between those of solid and vapor.
To set the notation, let $a$ be the triangular-lattice spacing and
$v_{\rm c}=(\sqrt{3}/4)a^2$ the elementary-cell volume -- the total number
of lattice sites is $M=V/v_{\rm c}$.
The $T=0$ values of $\Omega/M$ for the three phases are
$-\mu/9,(-3\epsilon-\mu)/12$, and 0, respectively.
Whence a liquid-vapor transition at $\mu=\mu_{\rm LV}\equiv -3\epsilon$,
followed by a solid-liquid transition at $\mu=\mu_{\rm SL}\equiv 9\epsilon$.
The coexistence pressures in reduced, $\epsilon/v_{\rm c}$ units are
$P_{\rm LV}^*=0$ and $P_{\rm SL}^*=1$, respectively.
A similar treatment for the LG34 model yields $T=0$ phase transitions at
$\mu=-3\epsilon$ and $\mu=4\epsilon$, with the same reduced values of the
coexistence pressures as for LG56.

In order to extend the analysis to non-zero (small) temperatures,
I proceed as follows. For each phase of the model, I list the allowed
configurations ({\it i.e.}, with no overlapping particles) that are
obtained from the ground state by removing or adding few particles.
Afterwards, for each excited state I calculate the multiplicity and
the Boltzmann weight in terms of $M,\beta\mu$, and $\beta\epsilon$
($\beta=(k_BT)^{-1}$, where $k_B$ is Boltzmann's constant).
Finally, for each pair of coexisting phases, particle configurations
are ordered by increasing relevance close to the transition threshold.
The goal is to come up with an approximate expression for the $\Omega$
of each phase that could next be used to draw the low-temperature
portion of the transition lines.

The detailed working out of these truncated expansions is rather lengthy.
In the Appendix, only a sketch of the derivation is presented, using the
LG56 model as an example. Besides that, also the relevant expansions for
the LG34 model are reported.
For both models, the solid-liquid $P_{\rm cox}(T)$ shows a decreasing
behavior at low temperature, leaving room for an IM in both cases
(see Figs.\,9-12).
While postponing to Section 4 the integration of these results with
the full transfer-matrix data, I here comment on the mechanism that is
responsible for a larger entropy of the solid phase at coexistence.
As specifically illustrated in the Appendix for the LG56 model, in the
liquid phase there is a strong imbalance between vacancies and
interstitials as for free-energy cost:
At low temperature, vacancies are far more easily excited than interstitials,
which instead require a considerable rearrangement of the local structure
in order to make room for the extra particles.
On the other hand, vacancies have a lower cost in the solid phase since
any particle removal in the liquid goes along with the rupture of many
nearest-neighbor bonds, hence with a substantial energy increase.
The outcome is that, at $T>0$, a smaller pressure than at
zero temperature is required for stabilizing the solid phase.

%
%
\section{Method}
\renewcommand{\theequation}{3.\arabic{equation}}
\setcounter{equation}{0}

To reconstruct the phase diagram of a lattice-gas model, I use the
transfer-matrix method and, to some extent, also grand-canonical
Monte Carlo (MC). If the interaction has an upper cutoff, the exact grand
potential of a system being infinite in one spatial direction and finite
in the other(s) can be computed as the logarithm of the maximum eigenvalue
$\lambda_{\rm max}$ of a so called transfer matrix (TM).
In two dimensions, the simplest case occurs when this matrix encodes the
interaction between a row of sites and the next row along the infinite
strip direction $y$.
In this case, the matrix size equals the total number of states in a row.
More generally, depending on the interaction range, the natural lattice unit
(NLU) can consist of just one single row, or a pair of consecutive rows,
or a triplet of rows, etc. In terms of $\lambda_{\rm max}$, the pressure
of the strip reads:
\be
P=\frac{k_BT}{{\cal N}_{\rm NLU}}\ln\lambda_{\rm max}\,,
\label{3-1}
\ee
${\cal N}_{\rm NLU}$ being the number of sites in the NLU.
The number density $\rho=(\partial P/\partial\mu)_T$ and its
$\beta\mu$ derivative, related to the isothermal compressibility
by $K_T=\rho^{-2}(\partial\rho/\partial\mu)_T$, are
evaluated from the raw $\beta P$ data as a three-point numerical first-
and second-order derivative, respectively ($\beta\mu$ is made to increase
by steps of 0.005). As a rule, the number of iterations that are necessary
to bring the maximum TM eigenvalue to convergence by the power method
is larger the larger $K_T$.
Upon increasing the number $N_x$ of sites in a row, phase-transition
signatures -- in the form of peaks in $(\partial\rho/\partial\mu)_T$ --
gradually emerge, allowing to extract the infinite-size behavior.
The virtue of the TM method is only limited by the range of the potential
and by its core extension, which both determine the maximum $x$ size that
can be stored in the computer.

The TM study of two-dimensional (2D) lattice gases received
a strong impulse in the mid sixties by the work of Runnels and
coworkers~\cite{Runnels2,Runnels3}: These authors
provided a considerable simplification in the problem by showing how to
reduce the TM size substantially without affecting $\lambda_{\rm max}$.
In practice, NLU states are grouped into equivalence classes bringing
together states that map onto each other when translated along $x$
and/or reflected with respect to the strip axis.
Whence, a matrix that is a condensed form of the TM is defined, of size
equal to the number of equivalence classes, whose maximum eigenvalue is
the same as for the original TM.
In Table 1, some typical dimensions are reported for the original TM and
for its compactified form, with reference to the LG34 and LG56 models.
Among many similar models, LG34 and LG56 provided the right compromise
between large core extension and feasibility of the TM analysis for
quite large sizes.

I have complemented the TM study with Metropolis MC simulations in the
grand-canonical ensemble. Typically, two million MC sweeps are generated
at equilibrium for $L\times L$ triangular lattices of increasing size,
up to a maximum of $L=240$, with periodic boundary conditions.
A MC sweep consists of one average attempt per site to change the occupation
number.
As a rule, simulation runs are carried out in a sequence, starting at a
low $\beta\mu$ value from the empty lattice and then raising $\beta\mu$
progressively at fixed temperature.
In the region of solid-liquid coexistence, equilibrium sampling could be
obstructed by liquid undercooling, {\it i.e.}, by hysteresis. This is why,
at high densities, the TM method (where available) is to be preferred to
MC simulations.

Among the computed quantities, the reduced number density
$\rho^*\equiv\rho\,v_{\rm c}$ and the isothermal compressibility $K_T$ are
especially monitored. These are expressed in terms of grand-canonical
averages as
\be
\rho^*=\left<{\cal N}\right>/L^2\,\,\,\,\,\,{\rm and}\,\,\,\,\,\,
\rho\,k_BTK_T=\frac{\left<\left(\Delta{\cal N}\right)^2\right>}{\left<{\cal N}\right>}\,,
\label{3-2}
\ee
with $\Delta{\cal N}={\cal N}-\left<{\cal N}\right>$, ${\cal N}$ being the
current particle number.

Finally, a useful tool for identifying the order of a phase transition
is to follow, e.g. for given $\beta$, the evolution of the MC density
histogram as a function of $\mu$. In a finite system, a roughly Gaussian
peak in this histogram is the imprint of a homogeneous phase,
a second-order transition (or just a crossover) is signalled by a broader
peak, while phase coexistence appears as a bimodal density distribution.
Hence, it is possible to discriminate between smooth/continuous and
first-order condensation by just looking at the $\mu$-evolution of
the density histogram at fixed temperature.

%
%
\section{Results}
\renewcommand{\theequation}{4.\arabic{equation}}
\setcounter{equation}{0}
\subsection{LG34 model}
This model was studied with the TM method only, considering just one
lattice strip 14 sites wide. This is the smallest size allowing both
solid ground states to be accommodated into the strip. The next suited
size, 28, is just too big for being amenable to numerical analysis by
the TM method.

In Fig.\,2, I show results for three isotherms somehow representative of
the different regimes, the overall phase diagram being represented in Fig.\,3.
It is clear that, beyond a fluid phase and a solid
of (reduced) density $\approx 1/4$, there is also an intermediate solid B
phase, of density $\approx 1/7$, which is present only at low temperature.
As temperature grows, the border which marks the region of solid B stability
gets thinner till it disappears, leaving a passage into the fluid.
However, this is an effect of the finite strip size while, in the
thermodynamic limit, there will likely be no way to enter solid B
smoothly from the fluid. Hence, calling ``solid'' the phase with
$\rho^*\approx 1/7$ (rather than ``liquid'') is a proper usage.
Finally, the fluid phase shows vapor-like properties at low pressure,
with densities lower than $1/7$, while appearing liquid-like for
high pressures, with densities higher than $1/7$.

The two solids and the liquid meet in a triple point at
$T^*\equiv k_BT/\epsilon\simeq 0.90$ and
$P^*\equiv Pv_{\rm c}/\epsilon\simeq 0.98$.
Above this pressure, and up to $P^*=1$, the solid-solid transition under
isobaric conditions anomalously occurs with volume contraction on heating,
as already anticipated from the low-temperature expansions.

Though the reported data are for just one strip size, their consistency
with an exact low-temperature analysis (see Figs.\,11 and 12) and the
sharpness of the transition signatures make me confident that the features
observed in Fig.\,3 give a faithful account of the phase behavior of
the model in the thermodynamic limit.

The LG34 model resembles the phase behavior of the purely repulsive
lattice-gas model studied in \cite{Runnels4}, hereby called LG3, where
particles exclude first- and second-neighbor sites on a triangular
lattice while softly repelling each other at third-neighbor distance.
Both models share the same phases, although LG34 is more effective in
promoting the stability of the solids at the expense of the fluid.
Moreover, the possibility of a solid-solid transition at constant
pressure, further accompanied by a density increase on heating,
is only peculiar to the LG34 model.

\subsection{LG56 model}
With respect to LG34, hard-core exclusion now comprises also third- and
fourth-neighbor sites. $N_x$ must be a multiple of 6 in order that both
solid ground states fit into the periodized strip. Fig.\,4 shows a
selection of results, relative to a number of isothermal paths for $N_x=18$.
At low temperature, there is a clear two-stage transition from vapor
to solid A, with an intermediate phase of density $\approx 1/12$.
This phase has strong crystalline features, as signalled by the
extremely small values of $K_T$, hence it will tentatively be called
solid B.

Upon increasing the temperature, the character of the first transition
gradually modifies, becoming smoother and smoother until it disappears
for $\beta\epsilon\approx 0.38$, leaving a direct transition from vapor
to solid A at higher temperatures. At the same time, the nature of the
intermediate phase also changes, since appreciable (0.1 or so) values of
$\partial\rho^*/\partial\beta\mu$ give support to the idea that this
phase is actually {\em liquid} (I shall add more evidence on this later).
The resulting phase diagram, see Fig.\,5, resembles that of $^4$He, with
the obvious difference that there is no superfluid region in the LG56
liquid.
The most exciting feature of the LG56 phase diagram is the clear IM
behavior that is found in the narrow pressure range from $\approx 0.97$
to 1 (see Fig.\,5 inset). The mechanism upon which IM rests has been
already clarified in Section 2: it deals with the lower cost of
proliferating vacancies in solid A than in the low-temperature solid B.

In the same Fig.\,5, I sketch the phase diagram of another model, called
LG5. Its pair potential shares the same core with $v_{56}(r)$, but it
shows a repulsive $\epsilon$ shoulder at $r=r_5$ and no well. This
purely-repulsive model cannot have a liquid phase, hence the
$\rho^*\approx 1/12$ phase should be a triangular solid which, in the
thermodynamic limit, will be separated from the fluid phase by an
uninterrupted first-order line.
The IM-like feature that is seen along the solid-solid coexistence line
of LG5 is probably an artefact of the finite strip size, since it deflates
upon going from $N_x=12$ to 18.

To probe the exactness of the TM technique, I have carried out a series
of MC simulations for a $18\times 360$ lattice at $\beta\epsilon=0.4$.
MC data points, reported as dots and asterisks in Fig.\,4, do clearly
lie superimposed over the TM data.
When the temperature is sufficiently high, a more significant check of the
TM results is provided by the calculation of the system pressure with the
method of thermodynamic integration. At high temperatures, the simulation
can be pushed through the transition region between liquid and solid, with
no risk of bumping into phase-space bottlenecks (in other words,
no hysteresis or effective ergodicity breaking is observed).
Eventually, $\beta\mu$ becomes so high that the truncated expansion
(\ref{A-10}) of the LG56 solid pressure holds true.
Combining this expansion with the MC values of the number density, I was
able to obtain an estimate of the pressure at the point of maximum of
$\partial\rho^*/\partial\beta\mu$ (the asterisk in Fig.\,5 inset) that
compares well with the TM datum.

The main question left open by the TM analysis concerns the nature of the
$\rho^*\approx 1/12$ phase: Is it actually a liquid? A related question is:
Does the liquid-vapor boundary of the infinite-size system really terminate
near $\beta\epsilon=0.38$ or it will rather join somewhere to the freezing
line, like in the LG5 case? Should the latter be true, it would indeed be
hard to qualify the intermediate phase of the LG56 model as liquid.

I have carried out MC simulations of large $L\times L$ triangular lattices,
up to $L=240$, along three isotherms, $\beta=030,0.34$, and 0.38. Each series
of runs was arrested at the edge of freezing, {\it i.e.}, before equilibrium
statistics could become unefficiently sampled by MC.
The values obtained for the number density and its $\beta\mu$ derivative are
reported in Fig.\,6, together with the TM data for $N_x=18$.
While leaving substantially unaltered (with respect to the TM hint) the
location of freezing, the MC results clearly indicate that
vapor condensation persists well beyond $T^*=1/0.38$, without merging into
the freezing transition (see the asterisks in Fig.\,7).

Further information on vapor condensation can be acquired by monitoring
the density histogram as a function of $\beta\mu$ at constant temperature.
This is drawn in Fig.\,7 inset for $L=240$ and $\beta\epsilon=0.38$
(similar results are found for $\beta\epsilon=0.34$ and 0.30).
Note the same Gaussian character and the comparable width of the density
histograms on either side of the condensation point.
While a first-order condensation can be safely excluded from these results,
it would be hard to discriminate a locus of critical points from a disorder
line. The numerical errors affecting particle-number fluctuations are not
small enough to decide, from the size scaling of the compressibility maximum,
whether the condensation line terminates with a critical ending point or
rather proceeds to infinite temperature.
In any case, a smooth vapor condensation can hardly be reconciled with
the symmetry breaking that would be implied by a phase transition into
a triangular solid.
It is true that there are lattice gases where the transition from vapor
to solid is reported to be second-order or even smoother, but this only
occurs when the core diameter of the particles is very small~\cite{Orban2}.
Moreover, the large values of $\partial\rho^*/\partial\beta\mu$ within the
$\beta\mu$ range of the intermediate phase are more appropriate to
a liquid than to a solid.

There is a last point to discuss, related to the possibility that a
phase-transition line separating solid B from liquid, running at
about constant $T$, was overlooked by the present TM study, which
considered only isothermal scans of the phase diagram.
It is worth noting that this was not the case for the LG34 model, where
in fact the TM analysis revealed the existence of a clear first-order
boundary between the two phases.
The question remains as to whether a smoother transition occurs in
the LG56 case.
To clarify this point, I carried out a TM study of the $N_x=12$ and 18
strips along various constant-$\mu$ lines. The locus of points where
$\partial\rho^*/\partial\beta\mu$ is maximum as a function of
temperature is reported as a dashed line in Fig.\,7.
By looking at this picture, one is tempted to conclude that solid B
and liquid are actually distinct phases.
In fact, things are more complicate since, for e.g. $\mu=8\epsilon$,
the broad maximum of $\partial\rho^*/\partial\beta\mu$ occurs with no
evidence of density jump -- see Fig.\,8.
My conclusion is that either this maximum marks the crossover from a
prominently liquid-like to a prominently solid B-like behavior within
the same phase or there is an underlying weak first-order transition
between liquid and solid B, characterized by a small jump of specific
volume. In the latter case, the seemingly negative slope of the
phase boundary would be the result, via Eq.\,(\ref{2-1}), of a slight
number-density decrease occurring upon going from liquid to solid B,
which is consistent with the TM data of Fig.\,8.

Further information is obtained from a series of MC simulations that I
carried out for $\mu=8\epsilon$, with $\beta\epsilon$ ranging from 0.3
to 0.4. In Fig.\,8, I report results for two sizes, $L=120$ and 240.
It turns out that the only clear singularity occurs at
$\beta\epsilon\simeq 0.34$, which corresponds to the same liquid-vapor
transition found at $\beta\mu\simeq 2.70$ along the $\beta\epsilon=0.34$
isotherm.
However, upon increasing $\beta\epsilon$ a little further, the simulated
system abruptly transformed into an almost perfect realization of solid B,
suggesting that what probably realizes in the LG56 model is the
weak-transition scenario: The low-temperature liquid, which looks like a
disordered patchwork of solid A and solid B grains, is not capable to
dismiss its solid A fraction continuously on cooling, being thus forced
to transform into solid B abruptly.
Considering that the hypothetical solid B-liquid boundary appears to join
to the locus of solid A melting at about where the slope of the latter
changes from negative to positive, the IM behavior of the LG56 model
might not differ significantly from the LG34 model, though the former is
undoubtedly closer to realize the ideal IM scenario than the latter.

Finally, let me draw some implications from the above results.
From the arguments presented in the Appendix, it is evident that
a decreasing freezing line on the $T$-$P$ plane will be the rule, at
sufficiently low temperature, for all the potentials having the same
shape of $v_{56}(r)$.
Whether this is an imprint of a genuine IM, rather than of a solid-solid
transition occurring with volume contraction upon heating at constant
pressure, is a complicate matter to grasp, which might be linked to the
existence of a congruous number of liquid configurations that are
proximal, as for number density, to the loose-packed crystalline ground
state. This is about to occur in the LG56 case, where the equilibrium
liquid at moderately low temperatures comes indeed very close, as for
specific volume and entropy contents, to a defected solid B.
Supposedly, when both particle core and potential well become slightly
enlarged with respect to the LG56 case, with the width of the barrier
staying fixed, the likelihood of observing a genuine IM will get enhanced.

%
%
\section{Conclusions}
Inverse melting (IM) is the phenomenon by which a crystal melts when
{\em cooled} at constant pressure. This can only occur if, at the
transition point, the solid is more entropic than the liquid.
Besides Helium at low temperature, the only system where an IM-like
transition is observed are some peculiar polymer solutions, denoted P4MP1.

In the present paper, I investigated the possibility of a lattice analog
of IM. To this aim, I introduced a triangular-lattice-gas system, called
LG56, where the particle diameter is 3, in units of the lattice constant,
and there is a narrow attractive well at a bit larger distance, $2\sqrt{3}$
lattice units.
The phase diagram of this system bears some indication of IM, similar to
that occurring in Helium by a completely different mechanism, which can
be rationalized as follows:
The soft inter-particle repulsion causes the existence of two distinct
crystalline ground states, a close-packed solid at high pressures and
a more open crystal structure at low pressures.
When the core diameter is large enough, which is about the case of the
LG56 model, the loose-packed crystal will smoothly transform into a
liquid at high temperatures, {\it i.e.}, without crossing a neat phase
boundary.
In addition to that, the interplay between interaction and lattice geometry
produces two effects, both essential to promote IM:
1) Interstitials are heavily suppressed in the low-temperature liquid;
2) vacancies are more easily excited in the solid than in the liquid
phase, thus conferring an entropic benefit to the solid.
As a result, the melting line on the $T$-$P$ plane bends downward at
sufficiently low temperature.

It can be argued that a genuine IM cannot be observed in a softly-repulsive
lattice-gas system since, upon lowering the temperature, any dense fluid
phase should eventually turn into a loose-packed solid.
In fact, the real question is whether a stable liquid can be pushed to
such low temperatures that, in a range of pressures, the close-packed
solid first melts into the liquid upon cooling, only after transforming
into the loose-packed solid. The present study shows that the LG56 model
is indeed close to realize this ideal IM scenario.

I add a final remark on the transferability of the above results to
soft-core potentials on a 3D lattice. As far as the previously cited
requisites on the inner-core and attractive-well extensions are met,
I find no reason to think that the behavior in 3D will be much different.
In particular, the same mechanism leading in 2D to a minimum in the freezing
pressure as a function of temperature will be at work also in 3D. Presumably,
the only marked difference concerns the order of the phase transitions,
which are stronger in a higher-dimensionality space. Consider, for instance,
a 3D counterpart of the LG56 model. While the freezing, solid A-to-liquid
transition of the LG56 model is strongly first-order already in 2D,
the transition from solid B to liquid would be much neater in 3D,
with the effect of removing any residual ambiguity on the nature of the
dashed-line crossover of Fig.\,7.

%
%
\begin{center}
{\bf Acknowledgements}
\end{center}
This work was supported by the University of Messina.
I wish to express my thanks to Paolo V. Giaquinta for introducing me to
the fascinating field of inverse melting and for a critical reading
of the manuscript. I am also grateful to Dora Magaudda at CECUM for
allowing me to use the ``Eneadi'' computer cluster prior to its full
implementation.

\newpage
%
%
\appendix
\section{Low-temperature expansions}
\renewcommand{\theequation}{A.\arabic{equation}}
\renewcommand{\thesubsection}{A.\arabic{subsection}}
\setcounter{equation}{0}

In this Appendix, exact series expansions are used to examine the
low-temperature behavior of the LG34 and LG56 models, with specific regard
to the determination of first-order phase boundaries.
Besides their intrinsic interest, these expansions may provide a consistency
check of the TM results.
For both lattice gases, an expansion of the grand partition function
$\Xi$, from which the grand potential $\Omega$ follows as $-k_BT\ln\Xi$,
is presented for all low-temperature phases: solid, liquid, and vapor
(in this appendix, I call ``liquid'' what is actually the loose-packed
solid B phase).
For a pair of competing phases, the transition line is located where
the respective grand potentials are equal.
The derivation of these expansions is rather lengthy and it would demand
too space to be reported here with full detail, hence it will be just
sketched. For one model (LG56), the high-$\mu$ expansion of the solid
pressure is also displayed.

\vspace{3mm}
LG56 model. --- In the very cold vapor, there are just a few particles on
an otherwise empty lattice. Leading terms in $\Xi$ are those associated
with the largest Boltzmann factors, hence with a small number of particles,
better if linked by liquid-like bonds. The expansion of $\Xi_{\rm V}$
will appear as:
\ba
\Xi_{\rm V} &=& 1+Me^{\beta\mu}+\frac{M(M-37)}{2}e^{2\beta\mu}+
3Me^{2\beta\mu}e^{\beta\epsilon}+3M(M-61)e^{3\beta\mu}e^{\beta\epsilon}+
9Me^{3\beta\mu}e^{2\beta\epsilon}
\nonumber \\
&+& 2Me^{3\beta\mu}e^{3\beta\epsilon}+
12Me^{4\beta\mu}e^{4\beta\epsilon}+3Me^{4\beta\mu}e^{5\beta\epsilon}+\ldots
\label{A-1}
\ea
The prefactors of the exponentials are multiplicities as calculated for the
bulk system.

Similarly, in the very cold liquid, the microstates which occur with
higher probability at equilibrium are obtained from the triangular crystal
of density $1/12$ by removing a small number of particles, better if bound
to each other.
The grand partition function starts with:
\ba
\Xi_{\rm L} &=& e^{(\beta\mu+3\beta\epsilon)M/12}\left[12+
Me^{-\beta\mu}e^{-6\beta\epsilon}+3Me^{-\beta\mu}e^{-11\beta\epsilon}+
3Me^{-2\beta\mu}e^{-11\beta\epsilon}\right.
\nonumber \\
&+& \frac{M}{2}\left(\frac{M}{12}-7\right)e^{-2\beta\mu}e^{-12\beta\epsilon}+
2Me^{-3\beta\mu}e^{-15\beta\epsilon}+9Me^{-3\beta\mu}e^{-16\beta\epsilon}
\nonumber \\
&+& 3M\left(\frac{M}{12}-10\right)e^{-3\beta\mu}e^{-17\beta\epsilon}+
3Me^{-4\beta\mu}e^{-19\beta\epsilon}+12Me^{-4\beta\mu}e^{-20\beta\epsilon}
\nonumber \\
&+& \left.6Me^{-5\beta\mu}e^{-23\beta\epsilon}+\ldots\right]\,.
\nonumber \\
\label{A-2}
\ea

Close to $\mu_{\rm LV}=-3\epsilon$ and $T=0$, I set
$\beta\mu=-3\beta\epsilon+\delta$. Moreover, I define the small parameter
$x=\exp({-\beta\epsilon})$. Disregarding all terms smaller than $x^8$, one
eventually finds:
\be
\frac{\ln\Xi_{\rm V}}{M}=e^{\delta}x^3+3e^{2\delta}x^5+\left( 2e^{3\delta}-
\frac{37}{2}e^{2\delta}\right) x^6
+\left( 3e^{4\delta}+9e^{3\delta}\right) x^7+\left( 12e^{4\delta}-
183e^{3\delta}\right) x^8+\ldots
\label{A-3}
\ee
and
\ba
\frac{\ln\Xi_{\rm L}}{M} &=& \frac{\delta}{12}+\frac{1}{12}e^{-\delta}x^3+
\frac{1}{4}e^{-2\delta}x^5+\left( -\frac{7}{24}e^{-2\delta}+\frac{1}{6}
e^{-3\delta}\right) x^6+\left( \frac{3}{4}e^{-3\delta}+
\frac{1}{4}e^{-4\delta}\right) x^7
\nonumber \\
&+& \left( \frac{1}{4}e^{-\delta}-\frac{5}{2}e^{-3\delta}+e^{-4\delta}+
\frac{1}{2}e^{-5\delta}\right) x^8+\ldots
\label{A-4}
\ea
In these formulae, we see the linked-cluster theorem at work: Despite
the fact that multiplicities of disconnected clusters of defects are
not linear in $M$, see (\ref{A-1}) and (\ref{A-2}), the logarithm of
the partition function turns out to be extensive due to cancellation
of terms proportional to $M^2,M^3$, etc.

For given $\beta$ and $\beta\mu$, the stable phase has the largest $\Xi$
or, equivalently, $\ln\Xi$. Stated differently, the vapor is more stable
than the liquid as far as $\Omega_{\rm V}<\Omega_{\rm L}$, or
\ba
\delta e^{\delta} &<& \left(12e^{2\delta}-1\right)x^3+
\left(36e^{3\delta}-3e^{-\delta}\right)x^5+
\left(24e^{4\delta}-222e^{3\delta}+\frac{7}{2}e^{-\delta}-
2e^{-2\delta}\right)x^6
\nonumber \\
&+& \left(36e^{5\delta}+108e^{4\delta}-9e^{-2\delta}-3e^{-3\delta}\right)x^7
\nonumber \\
&+& \left(144e^{5\delta}-2196e^{4\delta}-3+30e^{-2\delta}-12e^{-3\delta}-
6e^{-4\delta}\right)x^8+\ldots
\label{A-5}
\ea
In particular, for small $x$ and $\delta=0$ the stable phase is vapor,
meaning that the low-$T$ part of the liquid-vapor coexistence line is
bent upward in the $\mu$-$T$ plane, see a comparison with the TM data in
Fig.\,9 below.

As far as the solid is concerned, the low-cost excitations are defected
crystals with a small number of vacancies.
The expansion of the grand partition function starts with:
\be
\Xi_{\rm S}=e^{\beta\mu M/9}\left[9+Me^{-\beta\mu}+
\frac{M}{2}\left(\frac{M}{9}-1\right)e^{-2\beta\mu}+
2Me^{-2\beta\mu}e^{3\beta\epsilon}+
3Me^{-3\beta\mu}e^{7\beta\epsilon}+\ldots\right]\,.
\label{A-6}
\ee
For the liquid, I can still use Eq.\,(\ref{A-2}) since interstitials are
irrelevant at low temperature: Their accomodation requires such a huge
reorganization of the local liquid structure, along with the breaking of
so many bonds, that the loss in Boltzmann weight due to a higher energy
greatly overcomes the gain due to an increased particle number. Close to
$\mu_{\rm LS}=9\epsilon$ and $T=0$, I set $\beta\mu=9\beta\epsilon+\delta$
and $x=\exp({-\beta\epsilon})$, thus arriving at the following $\ln\Xi$
expansions for the liquid and the solid:
\be
\frac{\ln\Xi_{\rm L}}{M}=\beta\epsilon+\frac{\delta}{12}+
\frac{1}{12}e^{-\delta}x^{15}+\frac{1}{4}e^{-\delta}x^{20}+\ldots
\label{A-7}
\ee
and
\be
\frac{\ln\Xi_{\rm S}}{M}=\beta\epsilon+\frac{\delta}{9}+
\frac{1}{9}e^{-\delta}x^9+\frac{2}{9}e^{-2\delta}x^{15}-
\frac{1}{18}e^{-2\delta}x^{18}+\frac{1}{3}e^{-3\delta}x^{20}+\ldots
\label{A-8}
\ee
The solid is more stable than the liquid when $\Omega_{\rm S}<\Omega_{\rm L}$,
or
\be
\delta e^{\delta}>-4x^9+\left(3-8e^{-\delta}\right)x^{15}+
2e^{-\delta}x^{18}+\left(9-12e^{-2\delta}\right)x^{20}+\ldots
\label{A-9}
\ee
For small $x$, the above inequality is satisfied for $\delta=0$, indicating
that the solid-liquid coexistence line is bent downward at small temperature
(see Fig.\,9 above).
This holds true also on the $T$-$P$ plane, as one obtains from the values
of the liquid or solid $P^*=T^*\ln\Xi/M$ along the coexistence line
$\mu_{\rm cox}(T)$, see Fig.\,10 above.

Finally, the first few terms in the high-$\mu$ expansion of the solid
pressure for the LG56 model are:
\be
\beta Pv_c=\frac{\beta\mu}{9}+\frac{1}{9}e^{-\beta\mu}+
\frac{4e^{3\beta\epsilon}-1}{18}e^{-2\beta\mu}+
\left(\frac{1}{3}e^{7\beta\epsilon}-\frac{2}{3}e^{3\beta\epsilon}+
\frac{19}{27}\right)e^{-3\beta\mu}+\ldots
\label{A-10}
\ee
From here, the high-$\mu$ expansion of the number density follows
by differentiation.

\vspace{3mm}
LG34 model. --- Without entering much into details, which are similar as
for LG56, I list the expansions which are relevant for the low-temperature
analysis of the LG34 model.

The grand partition function of the vapor reads:
\ba
\Xi_{\rm V} &=& 1+Me^{\beta\mu}+\frac{M(M-25)}{2}e^{2\beta\mu}+
6Me^{2\beta\mu}e^{\beta\epsilon}+6M(M-40)e^{3\beta\mu}e^{\beta\epsilon}+
42Me^{3\beta\mu}e^{2\beta\epsilon}
\nonumber \\
&+& 4Me^{3\beta\mu}e^{3\beta\epsilon}+60Me^{4\beta\mu}e^{4\beta\epsilon}+
6Me^{4\beta\mu}e^{5\beta\epsilon}+\ldots
\label{A-11}
\ea

The grand partition function of the liquid (not to be confused with
the dense fluid phase of the model) reads:
\ba
\Xi_{\rm L} &=& e^{(\beta\mu+3\beta\epsilon)M/7}\left[14+
2Me^{-\beta\mu}e^{-6\beta\epsilon}+12Me^{-\beta\mu}e^{-10\beta\epsilon}+
6Me^{-2\beta\mu}e^{-11\beta\epsilon}\right.
\nonumber \\
&+& M\left(\frac{M}{7}-7\right)e^{-2\beta\mu}e^{-12\beta\epsilon}+
12Me^{-2\beta\mu}e^{-14\beta\epsilon}+4Me^{-3\beta\mu}e^{-15\beta\epsilon}+
18Me^{-3\beta\mu}e^{-16\beta\epsilon}
\nonumber \\
&+& 6M\left(\frac{M}{7}-10\right)e^{-3\beta\mu}e^{-17\beta\epsilon}+
6Me^{-4\beta\mu}e^{-19\beta\epsilon}+24Me^{-4\beta\mu}e^{-20\beta\epsilon}
\nonumber \\
&+& \left.12Me^{-5\beta\mu}e^{-23\beta\epsilon}+\ldots\right]\,.
\nonumber \\
\label{A-12}
\ea

With the same $x$ as before, the insertion of
$\beta\mu=-3\beta\epsilon+\delta$ into Eqs.\,(\ref{A-11}) and (\ref{A-12})
shows that the vapor is more stable than the liquid when
\ba
\delta e^{\delta} &<& \left(7e^{2\delta}-1\right)x^3+
\left(42e^{3\delta}-3e^{-\delta}\right)x^5+
\left(28e^{4\delta}-\frac{175}{2}e^{3\delta}+\frac{7}{2}e^{-\delta}-
2e^{-2\delta}\right)x^6
\nonumber \\
&+& \left(42e^{5\delta}+294e^{4\delta}-6-9e^{-2\delta}-3e^{-3\delta}\right)x^7
\nonumber \\
&+& \left(420e^{5\delta}-1680e^{4\delta}-6e^{-\delta}+30e^{-2\delta}-
12e^{-3\delta}-6e^{-4\delta}\right)x^8+\ldots
\label{A-13}
\ea
The ensuing liquid-vapor coexistence line on the $T$-$\mu$ plane is
reported in Fig.\,11 below.

The solid grand partition function reads:
\be
\Xi_{\rm S}=e^{\beta\mu M/4}\left[4+Me^{-\beta\mu}+
\frac{M}{2}\left(\frac{M}{4}-1\right)e^{-2\beta\mu}+
3Me^{-3\beta\mu}e^{4\beta\epsilon}+\ldots\right]\,.
\label{A-14}
\ee

Upon substituting $\beta\mu=4\beta\epsilon+\delta$ into Eqs.\,(\ref{A-12})
and (\ref{A-14}), it finally turns out that the solid is more stable than
the liquid when
\be
\delta e^{\delta}>-\frac{7}{3}x^4+\left(\frac{7}{6}e^{-\delta}-
7e^{-2\delta}\right)x^8+\frac{4}{3}x^{10}+\ldots
\label{A-15}
\ee
The solid-liquid coexistence line is reported in Fig.\,11 above.
Phase boundaries on the $T$-$P$ plane have a similar appearance, see Fig.\,12.

%
%

\newpage
%
%
\begin{center}
\large TABLE CAPTION
\normalsize
\end{center}
\begin{description}
\item[{\bf Table 1 :}]
Some data concerning the TM of the lattice-gas
models that are studied in this paper:
$N_x$ is the number of lattice sites along the finite strip size, chosen
in such a way as to comply with the periodicity of the two solid ground
states ({\it i.e.}, a multiple of 14 for LG34 and of 6 for LG56);
${\cal N}_{\rm NLU}$ gives the number of sites comprised in the natural
lattice unit ({\it i.e.}, $3N_x$ for LG34 and $4N_x$ for LG56); $M_1$ is
the size of the original TM; $M_2$ is the size of the symmetry-reduced
matrix, sharing the same leading eigenvalue with the original matrix.
\end{description}

\newpage
%
%
\begin{center}
\large FIGURE CAPTIONS
\normalsize
\end{center}
\begin{description}
\item[{\bf Fig.\,1 :}]
Pictorial description of the lattice-gas models
defined in Eq.\,(\ref{2-1}). 
When a particle is placed in the position marked as a full dot, crosses
denote forbidden sites for the centers of other particles. 
Open squares denote the lattice sites that are occupied by the
nearest-neighbor particles of the central particle in the close-packed
solid A.
Open dots denote attractive sites for the central particle: Here are
placed its nearest-neighbor particles in the loose-packed solid B.
In the LG34 model, these attractive sites are twelve, contributing a
twofold degeneracy to the spatial orientation of solid B.

\item[{\bf Fig.\,2 :}]
LG34 model, TM results for $N_x=14$.
The reduced number density (dotted lines) and its $\beta\mu$ derivative
(continuous lines) are shown for three isotherms, $\beta\epsilon=0.75,0.80$,
and $0.89$. To help the eye, straight lines are drawn between data points.
Peaks in the density derivative are the imprint of phase boundaries:
As a rule, the more pronounced a maximum is, the stronger
the phase transition. The confluence of three distinct phases (solid A,
solid B, and liquid) at $\beta\epsilon\approx 0.90$ is quite transparent
from the behavior at $\beta\epsilon=0.89$ (see also Fig.\,3 inset). Upon
raising the temperature, the region of solid B gradually shrinks until
its boundaries fade away. However, in the thermodynamic limit, there will
reasonably be no path to go smoothly from fluid to solid B.

\item[{\bf Fig.\,3 :}]
LG34 model, TM results for $N_x=14$.
Overall phase diagram of the system as resulting from joining the $(T,P)$
points that correspond to $\partial\rho^*/\partial\beta\mu$ peaks.
The LG34 potential (open dots and continuous lines) is contrasted with
the LG3 repulsive law of Ref.\,\cite{Runnels4} ($N_x=14$, open squares
and dashed lines). Spline interpolants drawn between LG34 data points
mark first-order phase boundaries; the interruptions are finite-size
effects. The fluid phase is vapor-like
at low pressure and liquid-like at high pressure. While both models can
exist in three phases (solid A, solid B, and fluid), LG34 is peculiar
in that it shows a solid-solid transition with volume contraction on
heating at constant pressure (inset).

\item[{\bf Fig.\,4 :}]
LG56 model, TM results for $N_x=18$.
The reduced number density (dotted lines) and its $\beta\mu$ derivative
(continuous lines) are shown for a number of isotherms, $\beta\epsilon=
0.34,0.38,0.4,0.45,0.5,0.6$, and $0.7$. To help the eye, straight lines
are drawn between data points. MC data for a $18\times 360$ periodic
sample, relative to $\beta\epsilon=0.4$, are also shown: Reduced number
density (open dots) and its $\beta\mu$ derivative (asterisks), as computed
through the number fluctuations -- see Eq.\,(\ref{3-2}).
Upon increasing the temperature, the liquid-vapor peak progressively
broadens till it disappears at $\beta\epsilon=0.38$.
The solid-liquid peak is present at all temperatures:
It is already very sharp for $\beta\epsilon=0.45$, becoming
even sharper for lower temperatures (not shown).

\item[{\bf Fig.\,5 :}]
LG56 model, TM phase diagram for $N_x=12$ (open
triangles and dashed lines) and $18$ (open dots and continuous lines).
The phase diagram of LG56 is contrasted with that of LG5 ($N_x=18$, open
squares and dashed lines).
Both models can exist in three phases: However, while the dense phase of
low pressure is solid B for the purely-repulsive LG5 model, the analogous
phase for LG56 is liquid, at least for high temperatures (see my arguments
in Section 4.B).
The LG56 model is peculiar in that an inverse melting occurs (inset).
The asterisk is a point on the solid-liquid coexistence line of the wider
LG56 strip that was constructed by MC (see main text).
The shallow minimum in the solid-solid $P_{\rm cox}(T)$ of LG5 is probably
a finite-size effect since this feature is less evident for $N_x=18$ than
for $N_x=12$ (not shown).

\item[{\bf Fig.\,6 :}]
LG56 model, data for three distinct isotherms,
$\beta=0.30,0.34$, and $0.38$. Above: reduced number density $\rho^*$;
below: $\beta\mu$ derivative of $\rho^*$.
TM data for $N_x=18$ (dotted lines) are contrasted with spline interpolants
of the MC results for $L\times L$ lattices, with $L=120,180$, and $240$
(continuous lines). The MC density derivative is computed through the
number fluctuations via Eq.\,(\ref{3-2}) (the larger $L$, the more
pronounced the $\partial\rho^*/\partial\beta\mu$ maximum is).
While distinctly recording the freezing transition at all temperatures,
a strip of 18 sites is not large enough to follow the boundary between
vapor and liquid beyond a reduced temperature of $1/0.38$. 
On the basis of the present MC evidence, it is hard to say whether the
liquid-vapor boundary survives at all temperatures or it rather turns
into a non-critical disorder line at a finite temperature (the scaling
of the MC compressibility at the estimated transition point is
inconclusive).

\item[{\bf Fig.\,7 :}]
LG56 model, phase diagram on the $T$-$\mu$ plane:
TM results for $N_x=18$ (open dots and continuous lines) plus three MC
data points for a $240\times 240$ lattice (asterisks), corresponding to
the location of the $\partial\rho^*/\partial\beta\mu$ maxima in Fig.\,6.
A further dashed line connects TM data points for $N_x=18$ recording
maxima of $\partial\rho^*/\partial\beta\mu$ along constant-$\mu$ cuts.
Inset: $\beta\mu$ evolution (from left to right) in the range from 2.30
to 3.25, with steps of 0.05, of the density histogram for $\beta=0.38$
and $L=240$.
It appears from the inset that the liquid-vapor transition either turns
continuous at high temperatures or it becomes a crossover.

\item[{\bf Fig.\,8 :}]
LG56 model, TM and MC data for the reduced number
density $\rho^*$ and its $\beta\mu$ derivative along the $\mu=8\epsilon$
line. TM data for $N_x=12$ (dashed line) and $N_x=18$ (dotted lines) are
contrasted with MC results for $L\times L$ lattices, with $L=120$ and $240$
(continuous lines).
The transition at $\beta\epsilon\simeq 0.34$ is from vapor to liquid while
a defected solid B structure abruptly appeared when pushing the simulation
beyond $\beta\epsilon=0.40$. 
Inset: $\beta\mu$ evolution (from left to right) in the range from 0.3
to 0.4, with steps of 0.005, of the density histogram for $L=240$.

\item[{\bf Fig.\,9 :}]
LG56 model, phase boundaries on the $T$-$\mu$ plane
as been obtained from the TM data ($N_x=12$: dashed lines; $N_x=18$: open
dots and continuous lines) and from exact low-temperature expansions (full
squares). Above: Solid-liquid coexistence (from left to right, squares refer
to $\delta=0,-0.01,-0.02,-0.05,-0.1,-0.2$); below: Liquid-vapor coexistence
(from left to right, squares refer to $\delta=0,0.01,0.02,0.05,0.1,0.15$).
As $\delta$ grows, the truncated expansions become less and less reliable
until consistency with TM data is lost.

\item[{\bf Fig.\,10 :}]
LG56 model, phase boundaries on the $T$-$P$ plane.
Data and notation as in Fig.\,9.

\item[{\bf Fig.\,11 :}]
LG34 model, phase boundaries on the $T$-$\mu$ plane
as been obtained from the TM data ($N_x=14$: open dots and continuous lines)
and from exact low-temperature expansions (full squares). Above: Solid-liquid
coexistence (from left to right, squares refer to
$\delta=0,-0.01,-0.02,-0.05,-0.1,-0.2$); below: Coexistence between liquid
and vapor (from left to right, squares refer to
$\delta=0,0.01,0.02,0.05,0.1,0.2,0.5$).
As $\delta$ grows, the truncated expansions become less and less reliable
until consistency with TM data is lost.

\item[{\bf Fig.\,12 :}]
LG34 model, phase boundaries on the $T$-$P$ plane.
Data and notation as in Fig.\,11.
\end{description}

\newpage
%
%
\begin{table}
\caption{\label{tab1}
}
\begin{tabular*}{\columnwidth}[c]{@{\extracolsep{\fill}}|c|r|r|r|r|}
\hline
model & $N_x$ & ${\cal N}_{\rm NLU}$ & $M_1$ & $M_2$\\
\hline\hline
LG34 & 14 & 42 & 397357 & 14715 \\
\hline\hline
LG56 & 6 & 24 & 88 & 16 \\
\hline
LG56 & 12 & 48 & 7768 & 385 \\
\hline
LG56 & 18 & 72 & 686905 & 19599 \\
\hline
\end{tabular*}
\end{table}

\newpage
%
%
\begin{figure}
\includegraphics[width=16cm,angle=0]{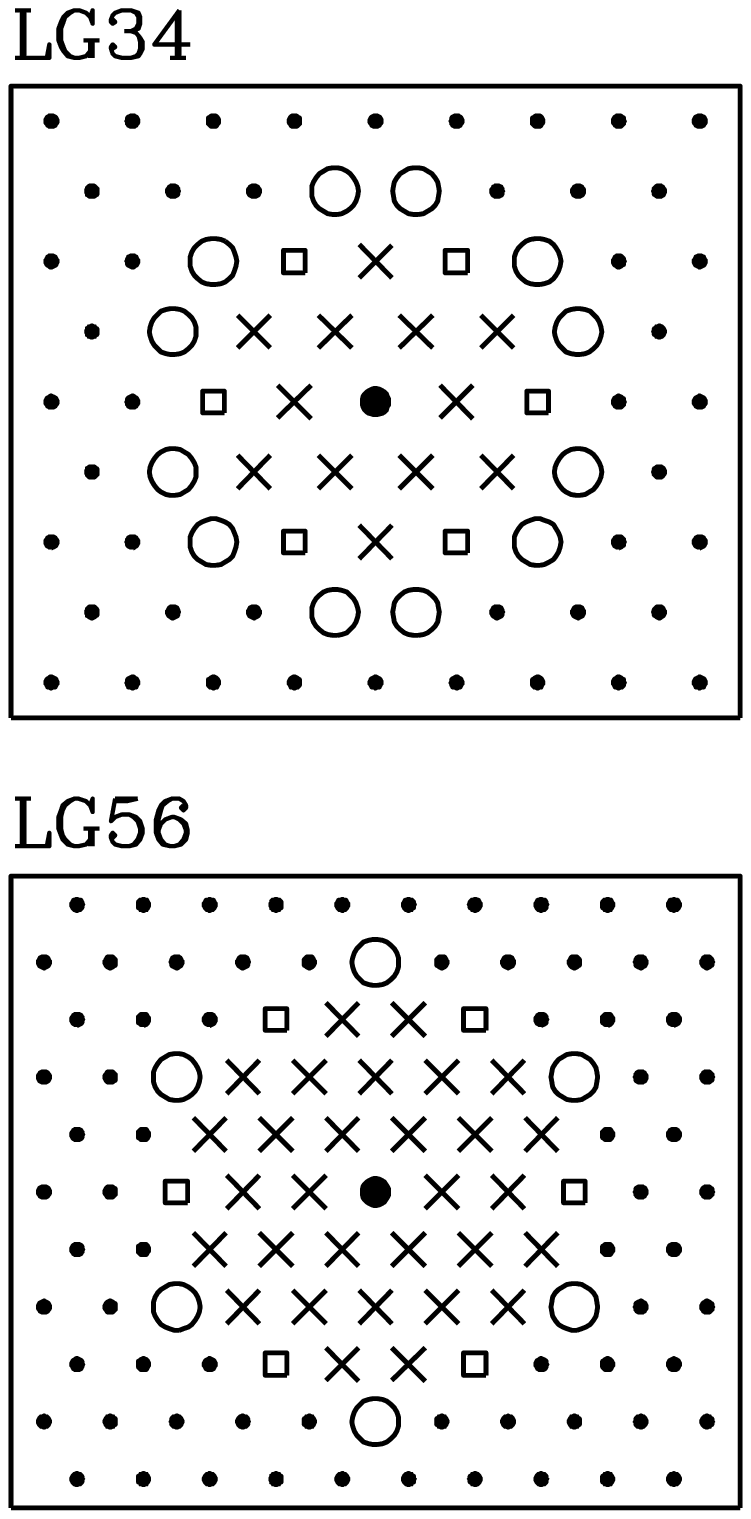}
\caption{\label{fig1}
Pictorial description of the lattice-gas models
defined in Eq.\,(\ref{2-1}).
When a particle is placed in the position marked as a full dot, crosses
denote forbidden sites for the centers of other particles.
Open squares denote the lattice sites that are occupied by the
nearest-neighbor particles of the central particle in the close-packed
solid A.
Open dots denote attractive sites for the central particle: Here are
placed its nearest-neighbor particles in the loose-packed solid B.
In the LG34 model, these attractive sites are twelve, contributing a
twofold degeneracy to the spatial orientation of solid B.
}
\end{figure}

\newpage
%
%
\begin{figure}
\includegraphics[width=16cm,angle=0]{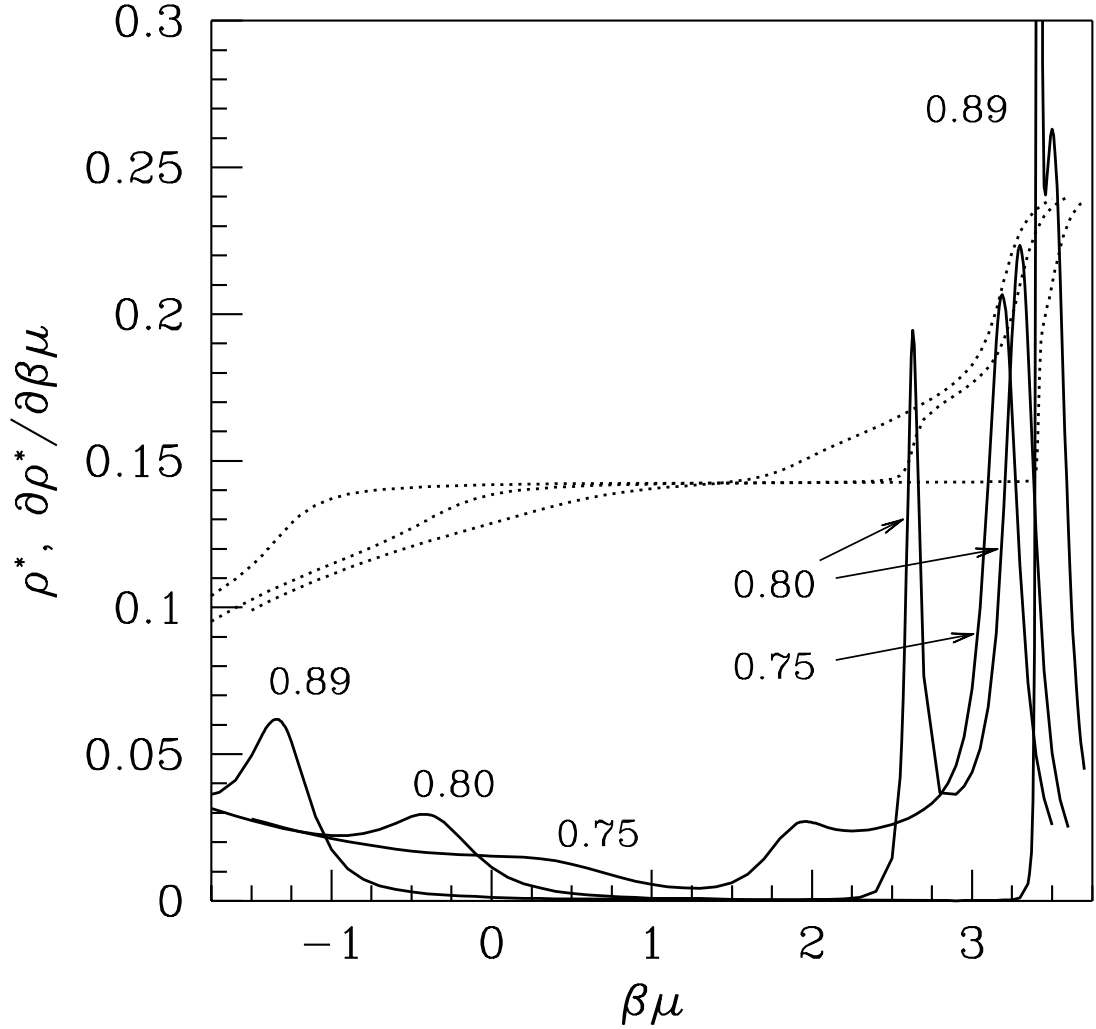}
\caption{\label{fig2}
LG34 model, TM results for $N_x=14$.
The reduced number density (dotted lines) and its $\beta\mu$ derivative
(continuous lines) are shown for three isotherms, $\beta\epsilon=0.75,0.80$,
and $0.89$. To help the eye, straight lines are drawn between data points.
Peaks in the density derivative are the imprint of phase boundaries:
As a rule, the more pronounced a maximum is, the stronger
the phase transition. The confluence of three distinct phases (solid A,
solid B, and liquid) at $\beta\epsilon\approx 0.90$ is quite transparent
from the behavior at $\beta\epsilon=0.89$ (see also Fig.\,3 inset). Upon
raising the temperature, the region of solid B gradually shrinks until
its boundaries fade away. However, in the thermodynamic limit, there will
reasonably be no path to go smoothly from fluid to solid B.
}
\end{figure}

\newpage
%
%
\begin{figure}
\includegraphics[width=16cm,angle=0]{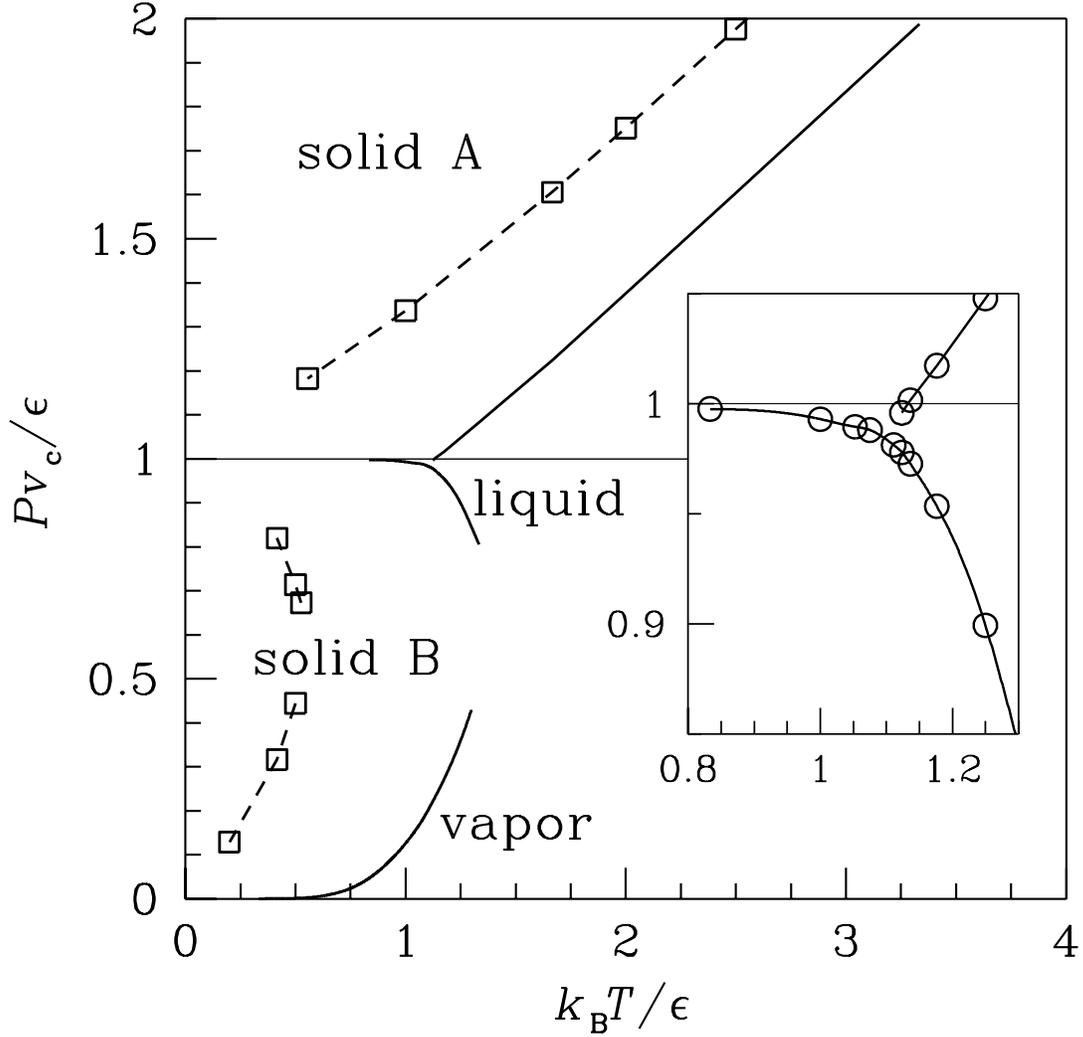}
\caption{\label{fig3}
LG34 model, TM results for $N_x=14$.
Overall phase diagram of the system as resulting from joining the $(T,P)$
points that correspond to $\partial\rho^*/\partial\beta\mu$ peaks.
The LG34 potential (open dots and continuous lines) is contrasted with
the LG3 repulsive law of Ref.\,\cite{Runnels4} ($N_x=14$, open squares
and dashed lines). Spline interpolants drawn between LG34 data points
mark first-order phase boundaries; the interruptions are finite-size
effects. The fluid phase is vapor-like
at low pressure and liquid-like at high pressure. While both models can
exist in three phases (solid A, solid B, and fluid), LG34 is peculiar
in that it shows a solid-solid transition with volume contraction on
heating at constant pressure (inset).
}
\end{figure}

\newpage
%
%
\begin{figure}
\includegraphics[width=16cm,angle=0]{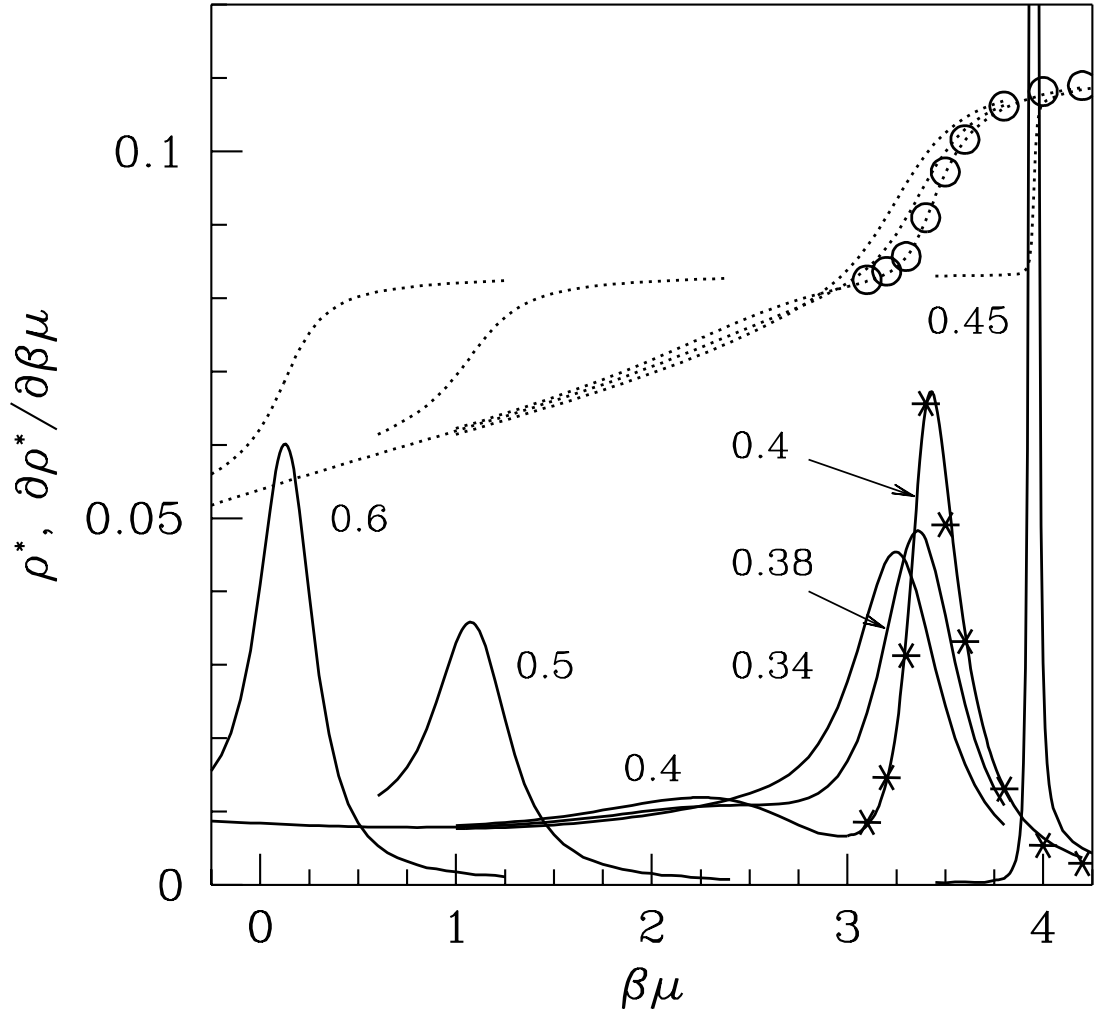}
\caption{\label{fig4}
LG56 model, TM results for $N_x=18$.
The reduced number density (dotted lines) and its $\beta\mu$ derivative
(continuous lines) are shown for a number of isotherms, $\beta\epsilon=
0.34,0.38,0.4,0.45,0.5,0.6$, and $0.7$. To help the eye, straight lines
are drawn between data points. MC data for a $18\times 360$ periodic
sample, relative to $\beta\epsilon=0.4$, are also shown: Reduced number
density (open dots) and its $\beta\mu$ derivative (asterisks), as computed
through the number fluctuations -- see Eq.\,(\ref{3-2}).
Upon increasing the temperature, the liquid-vapor peak progressively
broadens till it disappears at $\beta\epsilon=0.38$.
The solid-liquid peak is present at all temperatures:
It is already very sharp for $\beta\epsilon=0.45$, becoming
even sharper for lower temperatures (not shown).
}
\end{figure}

\newpage
%
%
\begin{figure}
\includegraphics[width=16cm,angle=0]{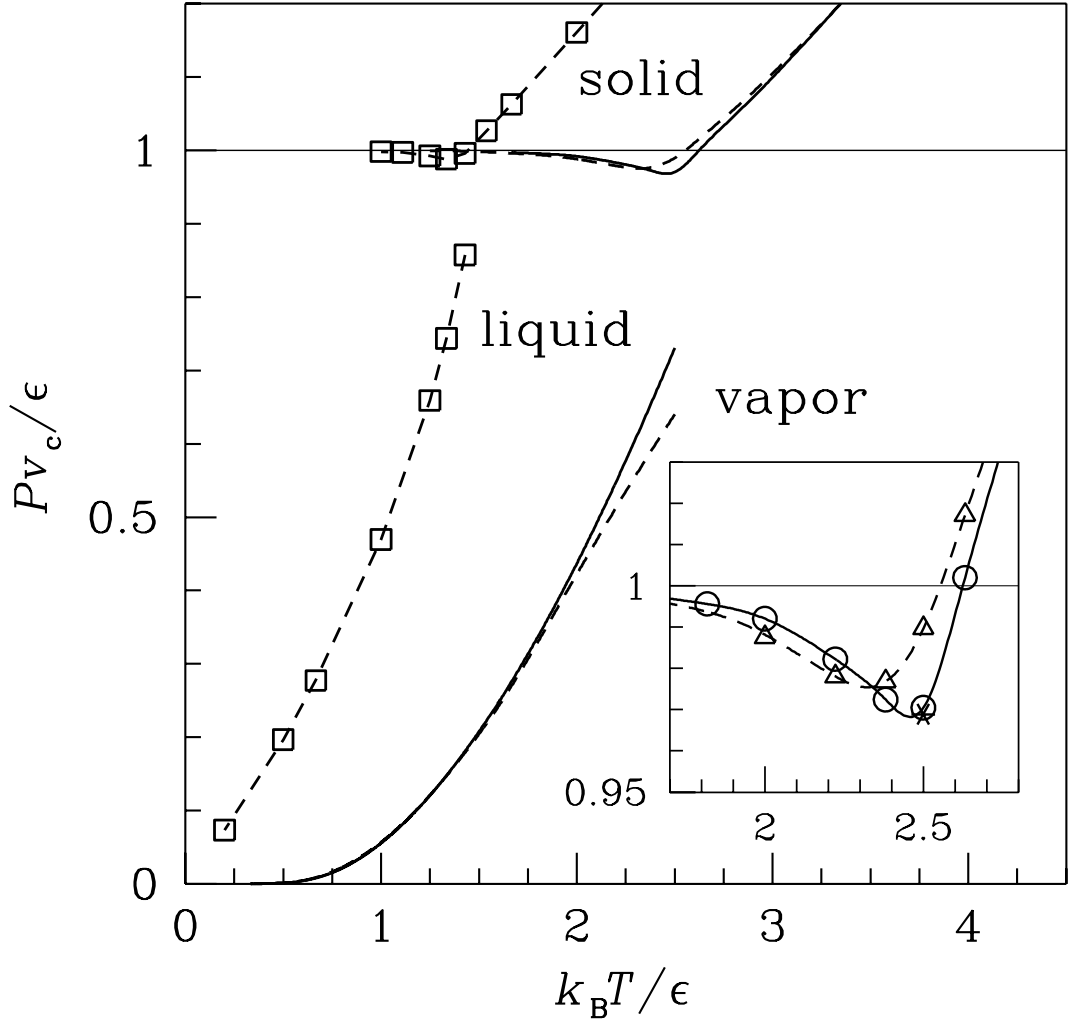}
\caption{\label{fig5}
LG56 model, TM phase diagram for $N_x=12$ (open
triangles and dashed lines) and $18$ (open dots and continuous lines).
The phase diagram of LG56 is contrasted with that of LG5 ($N_x=18$, open
squares and dashed lines).
Both models can exist in three phases: However, while the dense phase of
low pressure is solid B for the purely-repulsive LG5 model, the analogous
phase for LG56 is liquid, at least for high temperatures (see my arguments
in Section 4.B).
The LG56 model is peculiar in that an inverse melting occurs (inset).
The asterisk is a point on the solid-liquid coexistence line of the wider
LG56 strip that was constructed by MC (see main text).
The shallow minimum in the solid-solid $P_{\rm cox}(T)$ of LG5 is probably
a finite-size effect since this feature is less evident for $N_x=18$ than
for $N_x=12$ (not shown).
}
\end{figure}

\newpage
%
%
\begin{figure}
\includegraphics[width=16cm,angle=0]{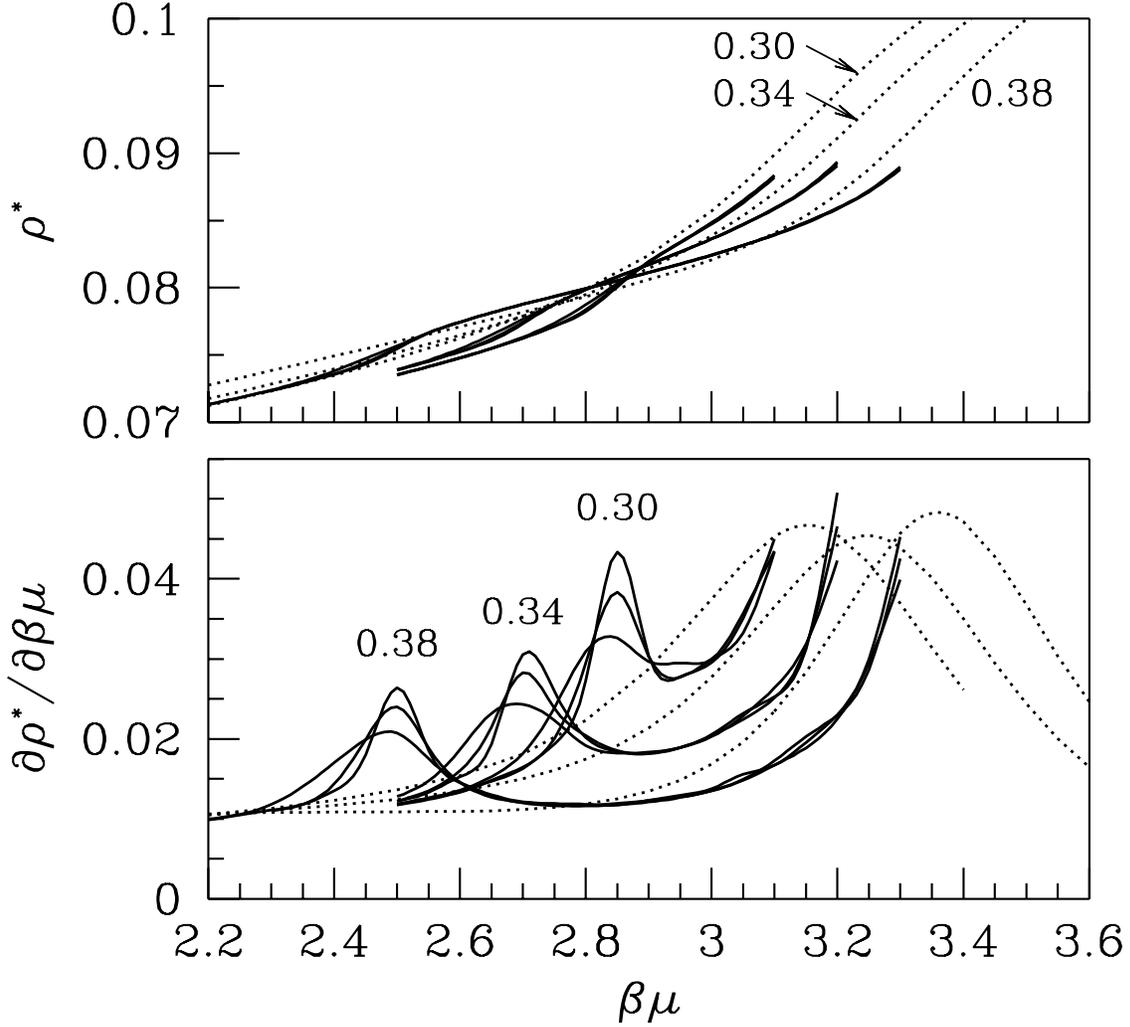}
\caption{\label{fig6}
LG56 model, data for three distinct isotherms,
$\beta=0.30,0.34$, and $0.38$. Above: reduced number density $\rho^*$;
below: $\beta\mu$ derivative of $\rho^*$.
TM data for $N_x=18$ (dotted lines) are contrasted with spline interpolants
of the MC results for $L\times L$ lattices, with $L=120,180$, and $240$
(continuous lines). The MC density derivative is computed through the
number fluctuations via Eq.\,(\ref{3-2}) (the larger $L$, the more
pronounced the $\partial\rho^*/\partial\beta\mu$ maximum is).
While distinctly recording the freezing transition at all temperatures,
a strip of 18 sites is not large enough to follow the boundary between
vapor and liquid beyond a reduced temperature of $1/0.38$.
On the basis of the present MC evidence, it is hard to say whether the
liquid-vapor boundary survives at all temperatures or it rather turns
into a non-critical disorder line at a finite temperature (the scaling
of the MC compressibility at the estimated transition point is
inconclusive).
}
\end{figure}

\newpage
%
%
\begin{figure}
\includegraphics[width=16cm,angle=0]{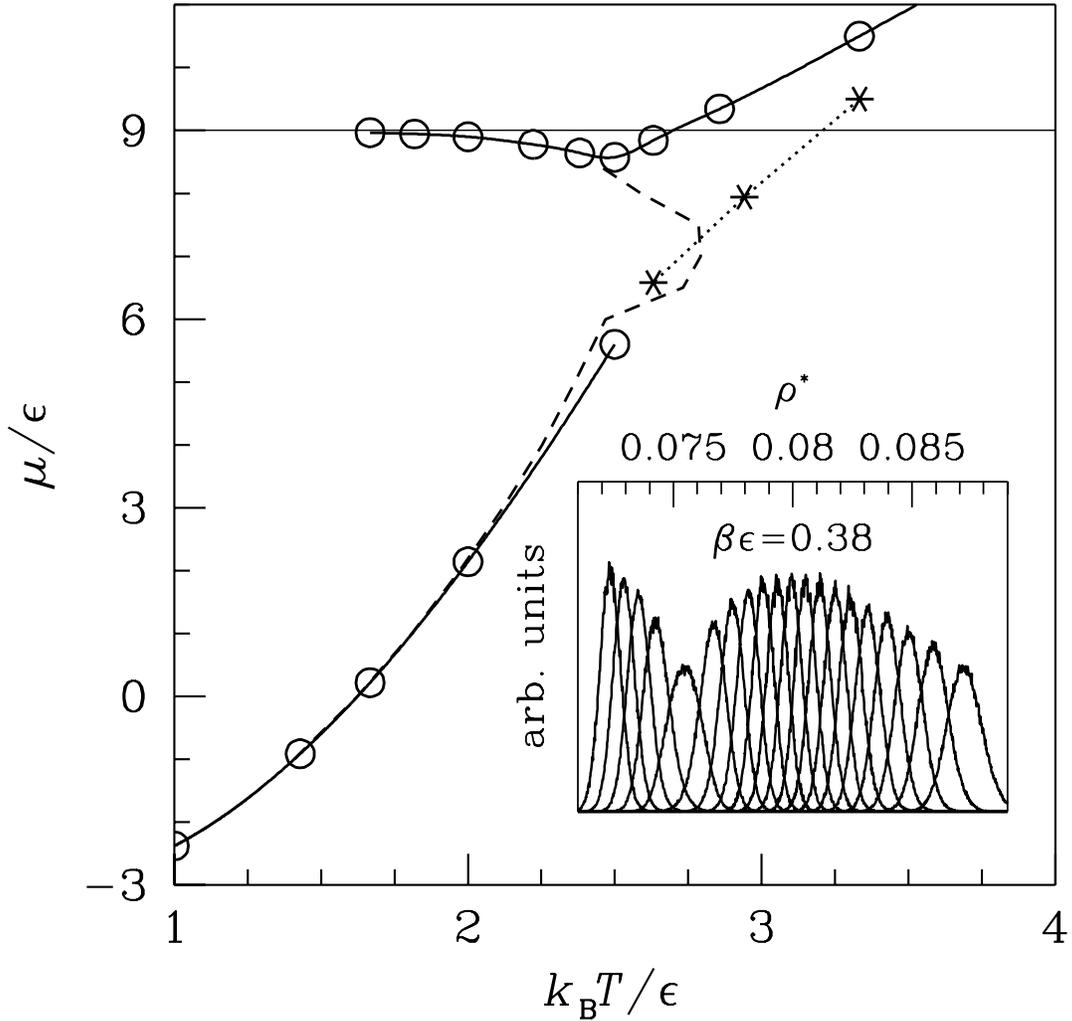}
\caption{\label{fig7}
LG56 model, phase diagram on the $T$-$\mu$ plane:
TM results for $N_x=18$ (open dots and continuous lines) plus three MC
data points for a $240\times 240$ lattice (asterisks), corresponding to
the location of the $\partial\rho^*/\partial\beta\mu$ maxima in Fig.\,6.
A further dashed line connects TM data points for $N_x=18$ recording
maxima of $\partial\rho^*/\partial\beta\mu$ along constant-$\mu$ cuts.
Inset: $\beta\mu$ evolution (from left to right) in the range from 2.30
to 3.25, with steps of 0.05, of the density histogram for $\beta=0.38$
and $L=240$.
It appears from the inset that the liquid-vapor transition either turns
continuous at high temperatures or it becomes a crossover.
}
\end{figure}

\newpage
%
%
\begin{figure}
\includegraphics[width=16cm,angle=0]{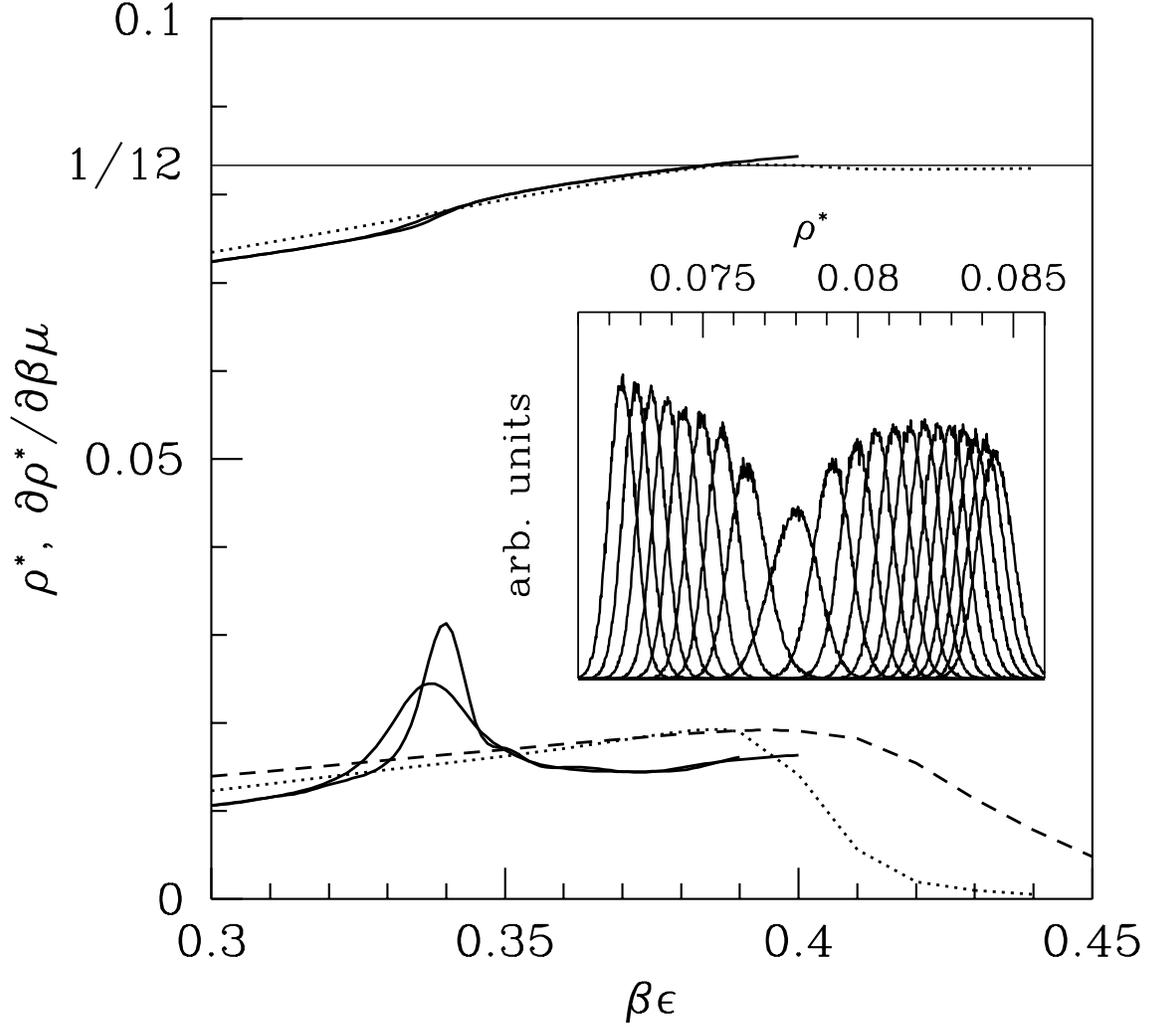}
\caption{\label{fig8}
LG56 model, TM and MC data for the reduced number
density $\rho^*$ and its $\beta\mu$ derivative along the $\mu=8\epsilon$
line. TM data for $N_x=12$ (dashed line) and $N_x=18$ (dotted lines) are
contrasted with MC results for $L\times L$ lattices, with $L=120$ and $240$
(continuous lines).
The transition at $\beta\epsilon\simeq 0.34$ is from vapor to liquid while
a defected solid B structure abruptly appeared when pushing the simulation
beyond $\beta\epsilon=0.40$.
Inset: $\beta\mu$ evolution (from left to right) in the range from 0.3
to 0.4, with steps of 0.005, of the density histogram for $L=240$.
}
\end{figure}

\newpage
%
%
\begin{figure}
\includegraphics[width=16cm,angle=0]{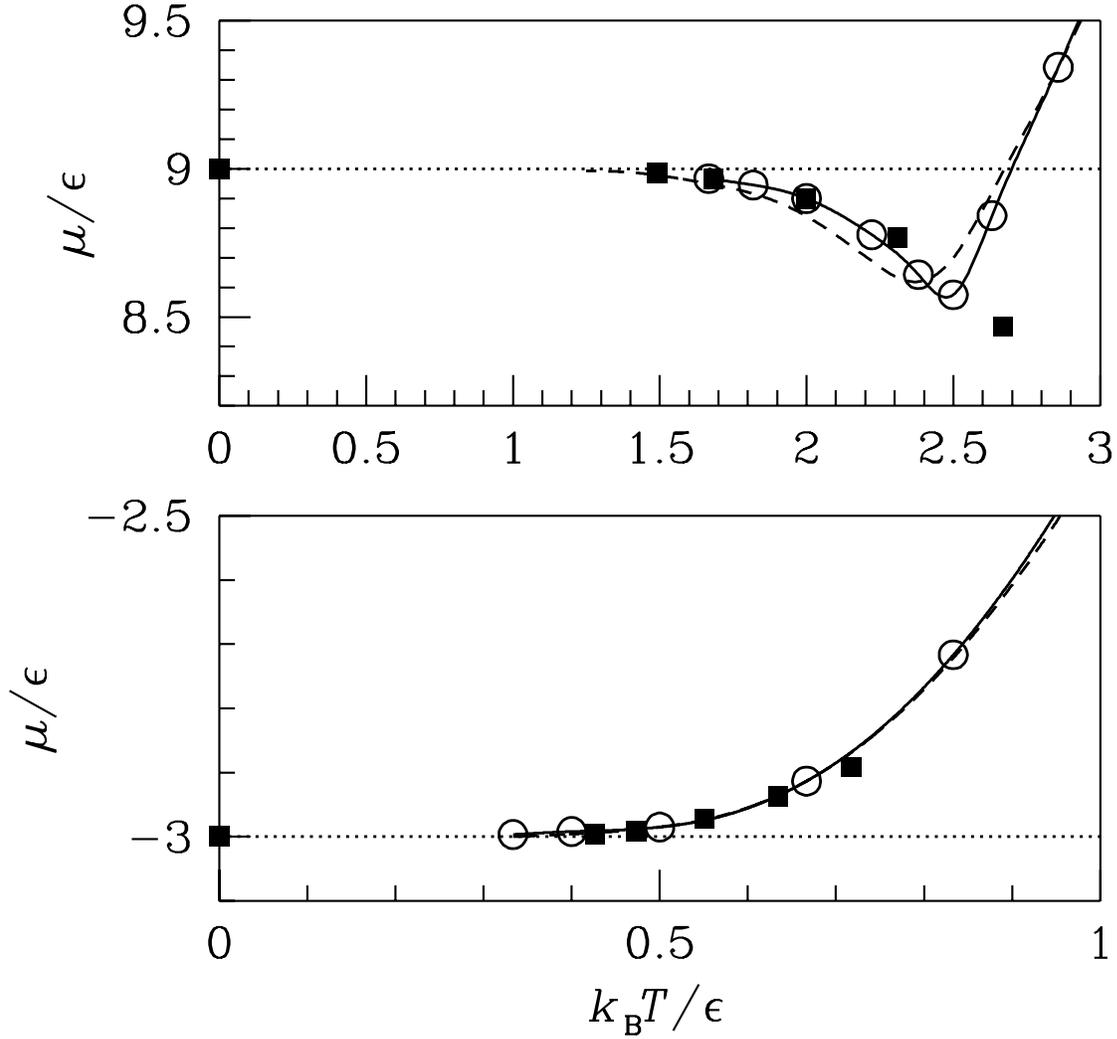}
\caption{\label{fig9}
LG56 model, phase boundaries on the $T$-$\mu$ plane
as been obtained from the TM data ($N_x=12$: dashed lines; $N_x=18$: open
dots and continuous lines) and from exact low-temperature expansions (full
squares). Above: Solid-liquid coexistence (from left to right, squares refer
to $\delta=0,-0.01,-0.02,-0.05,-0.1,-0.2$); below: Liquid-vapor coexistence
(from left to right, squares refer to $\delta=0,0.01,0.02,0.05,0.1,0.15$).
As $\delta$ grows, the truncated expansions become less and less reliable
until consistency with TM data is lost.
}
\end{figure}

\newpage
%
%
\begin{figure}
\includegraphics[width=16cm,angle=0]{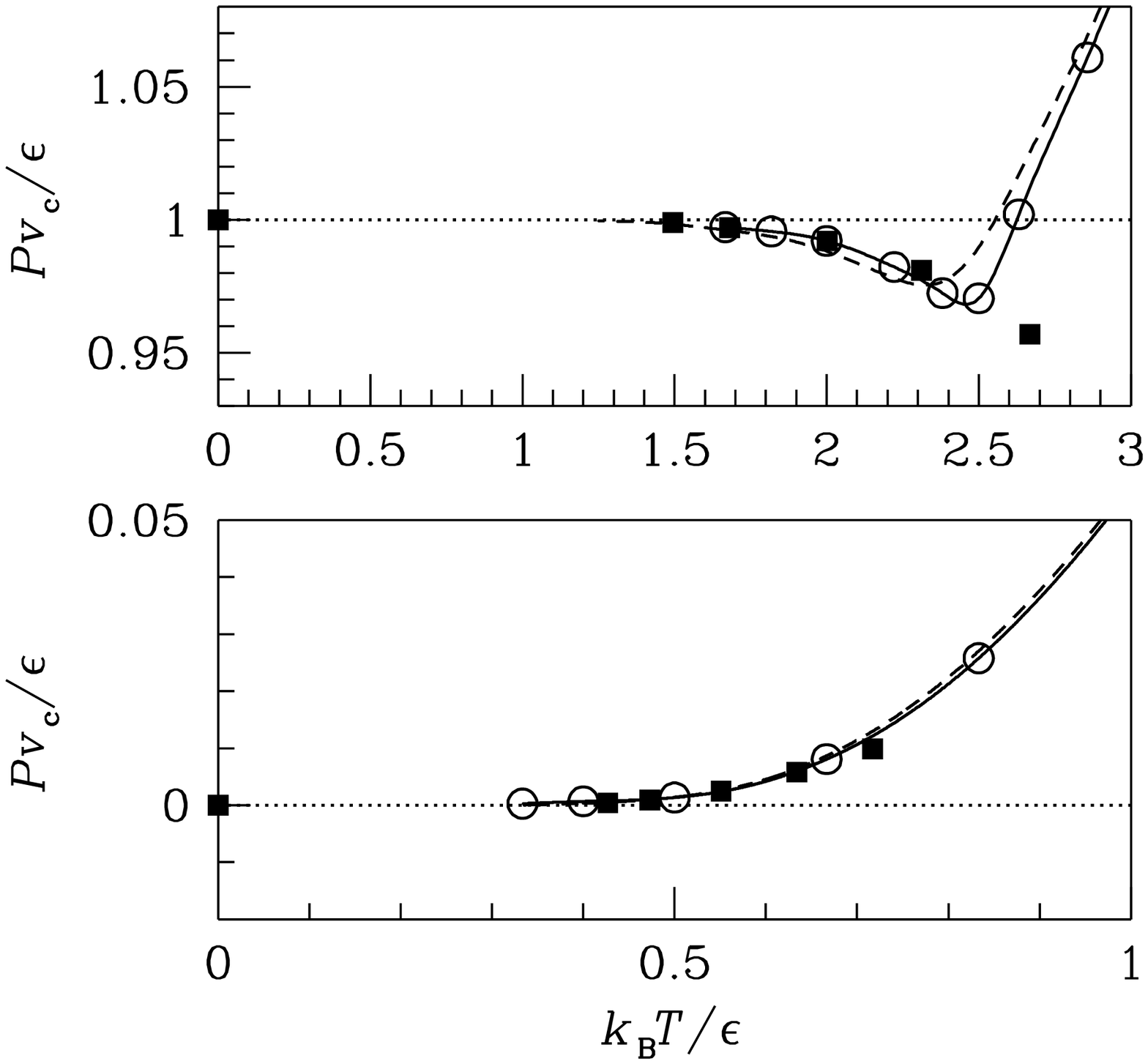}
\caption{\label{fig10}
LG56 model, phase boundaries on the $T$-$P$ plane.
Data and notation as in Fig.\,9.
}
\end{figure}

\newpage
%
%
\begin{figure}
\includegraphics[width=16cm,angle=0]{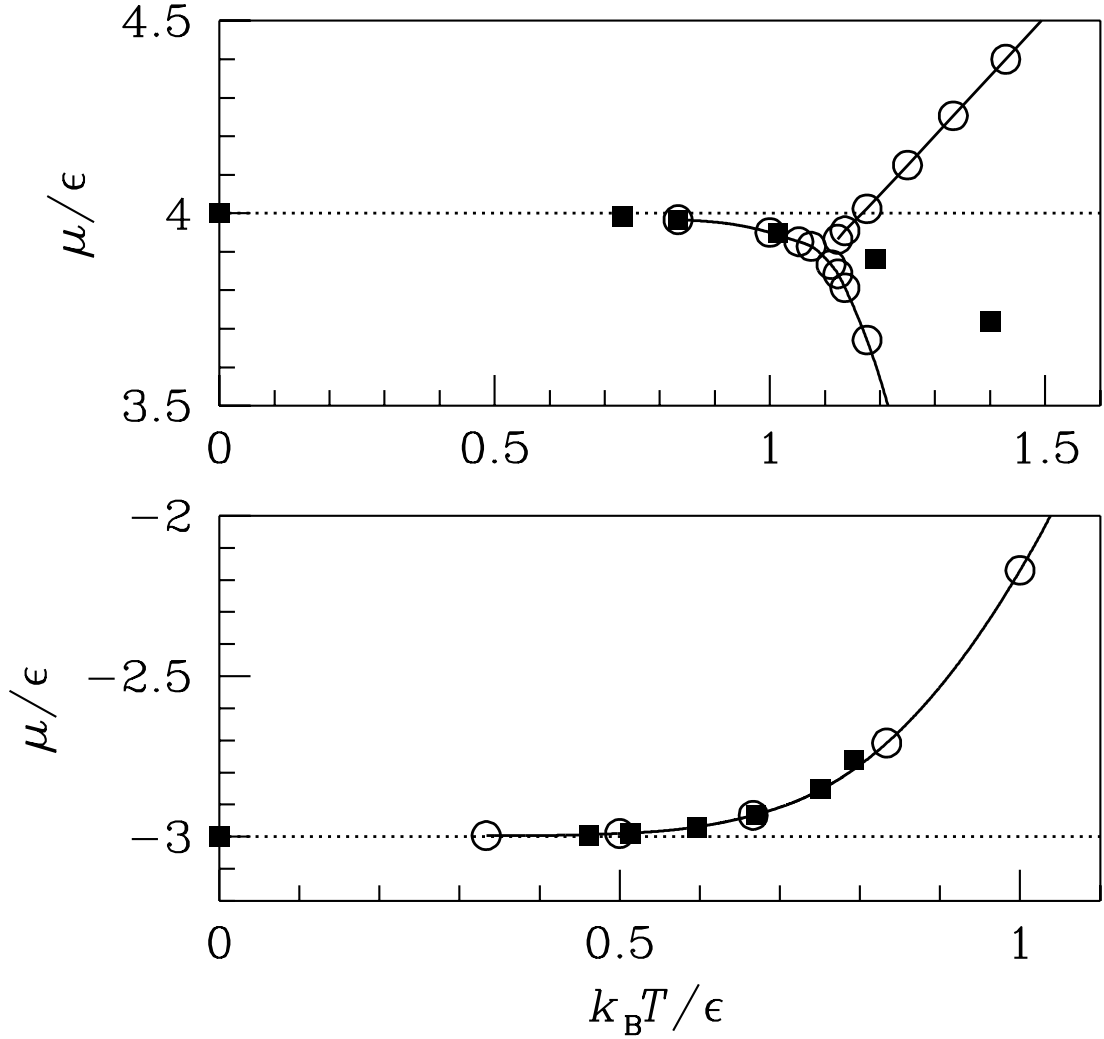}
\caption{\label{fig11}
LG34 model, phase boundaries on the $T$-$\mu$ plane
as been obtained from the TM data ($N_x=14$: open dots and continuous lines)
and from exact low-temperature expansions (full squares). Above: Solid-liquid
coexistence (from left to right, squares refer to
$\delta=0,-0.01,-0.02,-0.05,-0.1,-0.2$); below: Coexistence between liquid
and vapor (from left to right, squares refer to
$\delta=0,0.01,0.02,0.05,0.1,0.2,0.5$).
As $\delta$ grows, the truncated expansions become less and less reliable
until consistency with TM data is lost.
}
\end{figure}

\newpage
%
%
\begin{figure}
\includegraphics[width=16cm,angle=0]{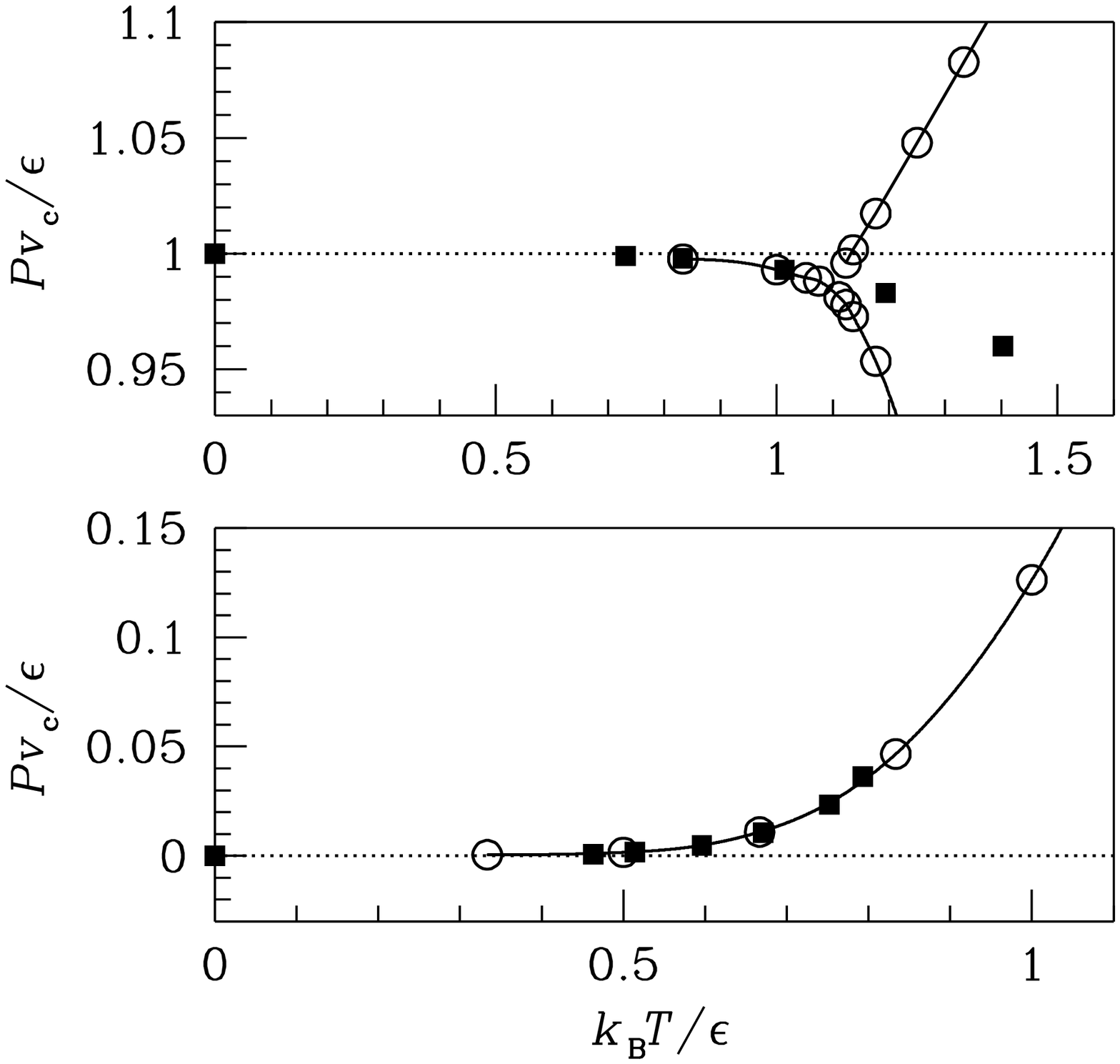}
\caption{\label{fig12}
LG34 model, phase boundaries on the $T$-$P$ plane.
Data and notation as in Fig.\,11.
}
\end{figure}
\end{document}